\documentclass[a4paper,aps,prd,10pt,twocolumn,superscriptaddress,nofootinbib,showpacs]{revtex4-1}
\usepackage{graphicx}
\usepackage{color}
\usepackage{amsmath}
\usepackage{amsfonts}
\usepackage{amssymb}
\usepackage{hyperref}
\usepackage{float}
\newcommand{\be}{\begin{equation}}
\newcommand{\ee}{\end{equation}}

\definecolor{purplerep}{rgb}{1,0.1,1}

\DeclareMathOperator{\sech}{sech}
\begin{document}

\title{Quantum matter bounce with a dark energy expanding phase}

\author{Samuel~Colin} \email{scolin@cbpf.br}
\author{Nelson~Pinto-Neto} \email{nelson.pinto@pq.cnpq.br}

\affiliation{Centro Brasileiro de Pesquisas F\'{\i}sicas,
Rua Dr.\ Xavier Sigaud 150 \\
22290-180, Rio de Janeiro -- RJ, Brasil}

\date{\today}

\begin{abstract}
Analyzing quantum cosmological scenarios containing one scalar field with exponential potential, we have obtained a universe model which realizes a classical dust contraction from very large scales, the initial repeller of the model, and moves to a stiff matter contraction near the singularity, which is avoided due to a quantum bounce. The universe is then launched in a stiff matter expanding phase, which then moves to a dark energy era, finally returning to the dust expanding phase, the final attractor of the model. Hence one has obtained a nonsingular cosmological model where a single scalar field can describe both the matter contracting phase of a bouncing model, necessary to give an almost scale invariant spectrum of scalar cosmological perturbations, and a transient expanding dark energy phase. As the universe is necessarily dust dominated in the far past, usual adiabatic vacuum initial conditions can be easily imposed in this era, avoiding the usual issues appearing when dark energy is considered in bouncing models.

\end{abstract}


\maketitle

\section{Introduction}

Scalar fields with exponential potentials have been widely studied in cosmology. Some arguments suggesting their origin from fundamental physics may be found, e.g. in Refs.~\cite{strings}. In an expanding (contracting) Friedmann model they possess attractor (repeller) solutions, where the pressure of the scalar field becomes proportional to its energy density ($p=w\rho$, $w=$const.). Such potentials have been used to model power-law inflation \cite{inflation}, dark energy \cite{darkenergy}, and the matter domination era of contracting phases of bouncing models \cite{bounce}, necessary for yielding an almost scale invariant spectrum of scalar cosmological perturbations in such models. The rich and interesting cosmological evolution dictated by such scalar fields in a flat Friedmann model in the framework of General Relativity was clearly described in Ref.~\cite{hewa}. For an expanding model, the universe emerges from a big bang singularity, where the scalar field behaves as stiff matter ($p=\rho$), and can either expand forever directly towards the attractor $p=w\rho$, or it can pass through a transient dark energy phase with $p\approx-\rho$, before reaching the attractor $p=w\rho$. In a contracting Friedmann model, the situation is time reversed: the universe contracts from the repeller $p=w\rho$ in the infinity past, and can either go directly to the singularity with $p=\rho$, or pass through a transient dark energy phase with $p\approx-\rho$ before reaching the singularity.

In this paper, we will investigate exponential potential scalar fields in the context of bouncing models, where the bounce happens due to quantum cosmological effects. In the context of Loop Quantum Cosmology (LQC), the bounce can be phenomenologically described through a modification of the Friedmann equation, $H^2 = 8\pi G/3 \rho(1-\rho/\rho_c)$, where $\rho_c$ is an energy density scale coming from the theory and very close to the Planck energy density \cite{lqc1}. The matter dominated contraction can be modeled by a scalar field with a suitable potential, which reduces to the exponential potential when $\rho\ll\rho_c$, where $\rho$ is the energy density of the scalar field \cite{lqc2} (see also Ref.~\cite{lqc3} for the case when an ekpyrotic phase is present). In this case, the universe realizes a classical matter dominated contraction until $\rho$ becomes comparable to $\rho_c$ and loop quantum effects realize the bounce, ejecting the universe in a classical matter dominated expansion when $\rho\ll\rho_c$ again. However, the richness of the exponential potential scalar field dynamics was not fully investigated in such descriptions. Here we will explore this richness in the framework of the de Broglie-Bohm quantum theory applied to the Dirac quantization of the classical minisuperspace model and its corresponding Wheeler-DeWitt equation. In this framework, the Wheeler-DeWitt equation is solved, and interpreted using the de Broglie-Bohm quantum theory (the usual Copenhaguen point of view cannot be used in quantum cosmology, see Ref.~\cite{fabris-nelson} for a review on this subject), where positions or field amplitudes (in the case of General Relativity, the spatial geometry) are assumed to have absolute reality. The quantum trajectories (the so called Bohmian trajectories) describing the scale factor evolution are calculated, and they are usually non-singular, presenting a bounce due to quantum effects at small scales, and turning to a classical standard evolution when the scale factor becomes sufficiently large, see Ref.~\cite{gaussian} for the case of a massless scalar field without potential (stiff matter with $p=\rho$). Of course the Wheeler-DeWitt quantization scheme cannot be viewed as the ultimate fundamental description of quantum gravitational effects, but it can be understood as a good approximation when the relevant physical scales are still some few orders of magnitude away from the Planck scale.

The Bohmian trajectories arising from the Wheeler-DeWitt quantization of Friedmann models with exponential potential scalar fields were never calculated. This is the aim of this paper. The space of solutions is explored in its full generality. There are solutions where the classical contracting phase is entirely dominated by dust, and quantum effects become relevant before the classical evolution leaves this phase, similar to the cases already studied in the LQC framework.
However, we also find wave solutions yielding Bohmian trajectories where the scalar field moves to a classical stiff matter contraction, directly or passing through a dark energy era
($p\approx -\rho$), as described in Ref.~\cite{hewa} for the classical dynamics, before quantum effects become relevant.
Then, in the quantum era, a bounce takes place, and the universe is ejected to an expanding phase of classical stiff matter domination, moving towards the matter expansion attractor ($p=0$), passing or not through a dark energy phase. Hence, such an exponential potential scalar field can, in a single shot, not only describe the matter contracting phase of a bouncing model, implying an almost scale invariant spectrum of scalar cosmological perturbations, but it can also model a transient dark energy era. Note that the presence of dark energy in the contracting phase of bouncing models turns problematic the imposition of vacuum initial conditions for cosmological perturbations in the far past of such models. For instance, if dark energy is a cosmological constant, all modes will eventually become bigger than the curvature scale in the far past, and an adiabatic vacuum prescription becomes quite contrived. Indeed, if we trace back in time this solution, it can alarmingly increase in the far past dominated by the cosmological constant. There are suggestions on ways to overcome this problem, but to our knowledge there is no net, consensual and clear solution to this issue, see Ref.~\cite{beatriz} and references therein for a discussion on this point. However, in the case of the scalar field with exponential potential, which contains a transient dark energy phase, the universe will always be dust dominated in the far past (running back in time, the dust repeller becomes an attractor), and adiabatic vacuum initial conditions can be easily imposed in this era. Hence, this is a situation where the presence of dark energy does not turn problematic the usual initial conditions prescription for cosmological perturbations in bouncing models.

Taking wave solutions of the Wheeler-DeWitt equation of the model, one can calculate their Bohmian trajectories and construct the space of solutions. Examining this space, one can easily see that the only possible bouncing solutions must have one and only one dark energy era, either in the contracting phase or in the expanding phase. Hence, the bounce solutions are necessarily asymmetric. This result must also be true for the phenomenological description of bounces in Loop Quantum Cosmology with such scalar fields, if analyzed in its full generality. Hence, taking the more realistic bouncing solutions where the dark energy phase happens in the expanding era, one has the picture of a universe which realizes a dust contraction from very large scales, the initial repeller of the model, moves to a stiff matter contraction near the singularity, realizes a quantum bounce which ejects the universe in a stiff matter expanding phase, which then moves to a dark energy era, finally returning to the dust expanding phase, the final attractor of the model.

In order to describe our results, the paper will be divided as follows: in section II, based on Ref.~\cite{hewa}, we summarize the classical minisuperspace model and its full space of solutions. In section III, we perform the Dirac quantization of the minisuperspace model obtaining the corresponding Wheeler-DeWitt equation, and we explain how the quantum trajectories can be obtained using the de Broglie-Bohm quantum theory. In section IV, we obtain wave solutions of the Wheeler-DeWitt equation, and we calculate their corresponding quantum trajectories, which describe universe models with the properties listed above. We end in section V with our conclusions and perspectives for future work.

\section{The classical minisuperspace model}

We are interested in a minisuperspace model described by the following action \cite{dks}
\begin{eqnarray}\label{action0}
S=\int L dt=\frac{3V}{\kappa^2}\int dt N\left(-\frac{a\dot{a}^2}{N^2}+K a-\frac{\Lambda a^3}{3}  \right)+\nonumber\\
\frac{1}{2}\int dt N a^3\left(\frac{\dot{\phi}^2}{N^2}-2 V_0 e^{-\kappa\lambda\phi}\right)\,,
\end{eqnarray}
where we set $\hbar = c=1$, $\kappa^2=8\pi G = 8\pi l^2_{\rm Pl}$, $N$ is the lapse function, $\Lambda$ is the cosmological constant, $K$ is a curvature index, $\lambda$ is a dimensionless coupling constant, and $V$ is the comoving volume of the spacelike homogeneous hypersurfaces, which we can set to be $V=4\pi l_{\rm Pl}^3 /3$, implying that when $a=1$ the volume of these surfaces is the Planck volume. There are two degrees of freedom:
$a(t)$, the scale factor of the Friedmann universe and $\phi(t)$, the homogeneous scalar field.
In what follows, we will assume $N=1$, $\Lambda=0$, $K=0$. As we will see later on, the space of solutions for such models possess an attractor and repeller where the pressure $p$ and energy density $\rho$ of the scalar field satisfies $p=w\rho$, $w=$const, where $\lambda^2 = 3(1+w)$. As we are interested in models with a long dust contraction, we choose $\lambda=\sqrt{3}$.

\begin{figure}[H]
\begin{center}
\includegraphics[width=0.4\textwidth]{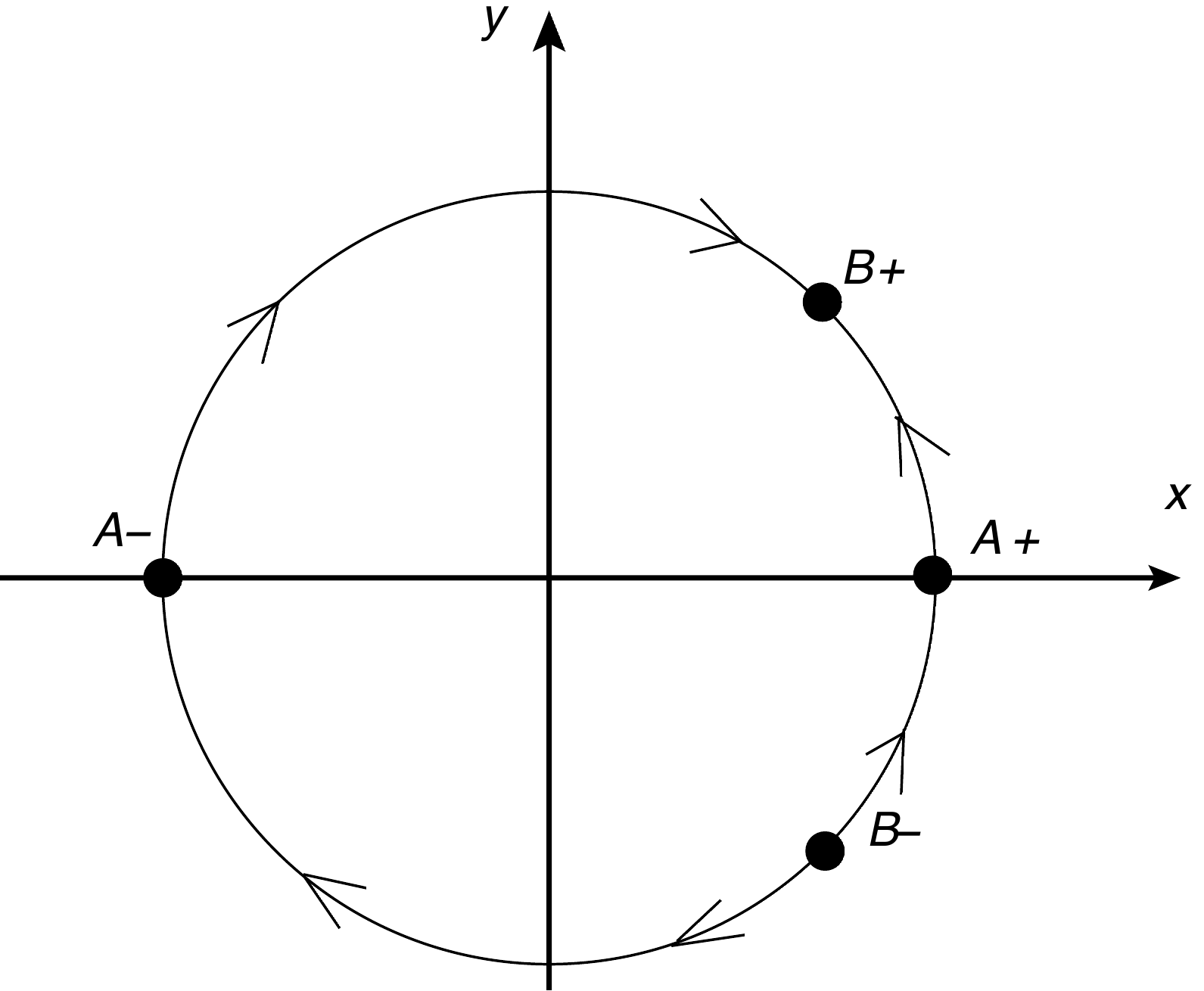}
\end{center}
\caption{Phase-space picture of the classical model in the variables $x$ and $y$ (see (\ref{hewaxy}) and Ref.~\cite{hewa})).
Arrows indicate evolution in cosmic time.}
\label{fighewa}
\end{figure}

In Ref.~\cite{hewa}, it was shown that the classical motion (for $\alpha =\ln a$ and $\phi$) takes place on a circle of radius $1$ (see Fig. (\ref{fighewa})) if one introduces the dimensionless variables $x$ and $y$, defined by
\be\label{hewaxy}
x=\frac{\kappa\dot{\phi}}{\sqrt{6}H}\,,\quad y=\frac{\kappa\sqrt{|V|}}{\sqrt{3}H}~,
\ee
where $H$ is the Hubble parameter ($H=\dot{a}/a=\dot{\alpha}$). Introducing the dimensionless scalar
field ${\bar{\phi}} = \kappa \phi/\sqrt{6}$ and omitting the bars, one has $V=V_0 e^{-3\sqrt{2}\phi}$.

The equations for $x$ and $y$ are

\begin{equation}
\label{x}
\frac{d x}{d \alpha} = - 3x(1-x^2)+\frac{3}{\sqrt{2}} y^2  ,
\end{equation}

\begin{equation}
\label{y}
\frac{d x}{d \alpha} = xy\biggl(3x - \frac{3}{\sqrt{2}}\biggr)  ,
\end{equation}
subjected to the constraint

\begin{equation}
\label{constraint}
x^2+y^2 = 1 .
\end{equation}
The ratio $p/\rho$ reads,

\begin{equation}
\label{ratio}
\frac{p}{\rho} = \frac{\dot{\phi}^2/2-V}{\dot{\phi}^2/2+V} = \frac{x^2 - y^2}{x^2 + y^2} .
\end{equation}

\begin{figure}[H]
\begin{center}
\includegraphics[width=0.4\textwidth]{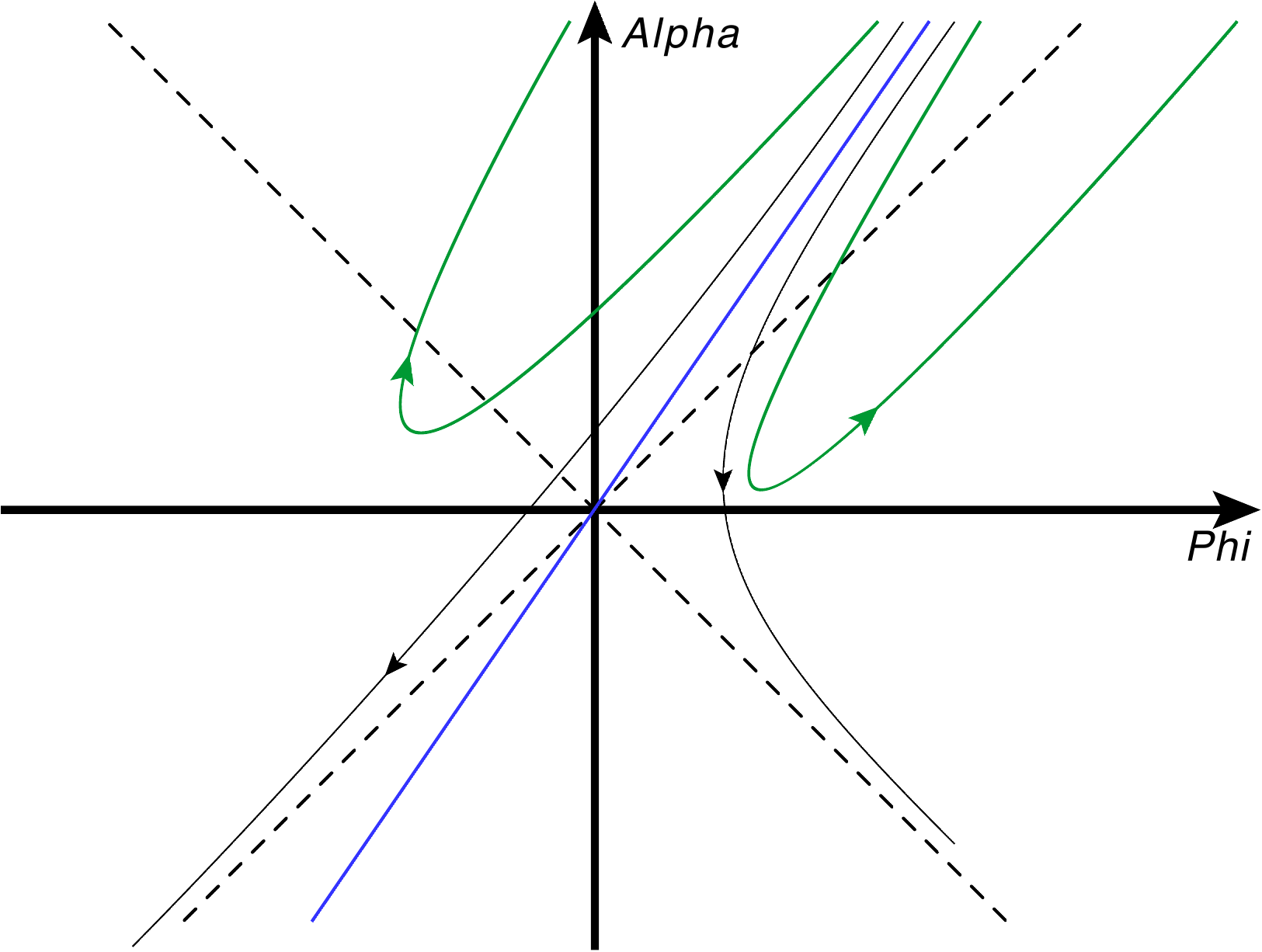}
\end{center}
\caption{Figure presenting classical trajectories (continuous lines with arrows indicating the flow of time), and hypothetical bouncing trajectories
(green continuous lines with arrows and $\alpha$ positive) in the $\alpha-\phi$ plane. The dotted lines correspond to $\alpha=\pm\phi$ (stiff matter behavior), whereas the continuous line (without arrow) corresponds to the classical attractor
$\alpha=\sqrt{2}\phi$ (dust behavior). The dark energy era happens when trajectories have vertical inclination.}
\label{figscan}
\end{figure}

The one-dimensional phase-space picture of these solutions is shown in Fig. (\ref{fighewa}), see Ref.~\cite{hewa} for details. The upper half-circle corresponds to an expanding universe ($y>0$) whereas the lower one corresponds to a contracting universe. The points with $y=0$, that is $A^{+}$ and $A^{-}$,
correspond to the singularity $a=0$. From Eq.~\eqref{ratio}, one can see that near this point the scalar field behaves as stiff matter, and $x=\pm 1 \Rightarrow \alpha = \pm \phi +$const., respectively.
The points $B^{+}$ and $B^{-}$ are the attractor and repeller, respectively. As $y=\pm x = 1/\sqrt{2}$ at these points, one can see from Eq.~\eqref{ratio} that the scalar field behaves like dust, and $x=1/\sqrt{2} \Rightarrow \alpha = \sqrt{2} \phi +$const.. The dark energy era ($p\approx -\rho$) happens around the region $x=0 \Rightarrow \phi =$const. (see again Eq.~\eqref{ratio}).
The classical trajectories for contracting universes, which start in the vicinity of $B^{-}$, can also be visualized in the $(\alpha,\phi)$ plane (see Fig. (\ref{figscan})).
In the present work, we would like to know if quantum effects allow the existence of
bouncing trajectories (these hypothetical bouncing trajectories are also shown in Fig. (\ref{figscan})).
From this figure one can guess that all bouncing trajectories with classical limit for large scale factors must have vertical inclination only once, which means that they must have a dark energy phase either in the contracting phase
or in the expanding phase.
On the circle of radius $1$ of Fig.~$1$, these trajectories would correspond to the following possible evolutions:
\be
\textrm{Bounce I}: B_{-}\rightarrow A_{+}\rightarrow A_{-}\rightarrow B_{+}
\ee
where the dark energy phase occurs in the expanding era, or
\be
\textrm{Bounce II}: B_{-}\rightarrow A_{-}\rightarrow A_{+}\rightarrow B_{+}~.
\ee
where the dark energy phase happens in the contracting era. In the following sections, we will exhibit explicit bouncing solutions with these behaviors.

\section{Quantization of the minisuperspace model}

We will now Dirac quantize the classical minisuperspace model and obtain its corresponding Wheeler-DeWitt equation.
The action (\ref{action0}) (with $K=\Lambda=0$, and $\lambda=\sqrt{3}$) reads,
\be\label{action}
S=\int L dt=\frac{1}{2}\int dt N e^{3\alpha}\biggl(-\frac{{\dot{\alpha}}^2}{N^2}+\frac{{\dot{\phi}}^2}{N^2}-2V_0 e^{-3\sqrt{2}\phi}\biggr)~,
\ee
where introduced the dimensionless time ${\bar{t}} = \kappa^2 t/(6 V)=t/ t_{\rm{Pl}}$, and a dimensionless potential amplitude,
${\bar{V}}_0 = 4\pi l_{\rm Pl} ^4 V_0/3$, and we have omitted the bars.
The momenta conjugate to $\alpha$ and $\phi$ are,
\be
\pi_{\alpha}=\frac{\partial L}{\partial\dot{\alpha}}\equiv m_\alpha\dot{\alpha}=-\frac{e^{3\alpha}}{N}\dot{\alpha}\,,\quad
\pi_{\phi}=\frac{\partial L}{\partial\dot{\phi}}\equiv m_\phi\dot{\phi}=\frac{e^{3\alpha}}{N}\dot{\phi} ,
\ee
yielding the Hamiltonian
\begin{align}
H=\pi_{\alpha}\dot{\alpha}+\pi_{\phi}\dot{\phi}-L\equiv N{\cal{H}}=\nonumber\\
\frac{N}{2}e^{-3\alpha}\left(-\pi_\alpha^2+\pi^2_\phi+2V_0 e^{6\alpha-3\sqrt{2}\phi}\right)~.
\end{align}
The Dirac quantization procedure imposes that the physical quantum state should be annihilated by the constraint operator, $\widehat{{\cal{H}}}\Psi=0$, yielding the Wheeler-DeWitt (WDW) equation,
\be
\left\{-\frac{\hat{\pi}_\alpha^2}{2}+\frac{\hat{\pi}_\phi^2}{2}+V_0 e^{6\alpha-3\sqrt{2}\phi}   \right\}\Psi(\alpha,\phi)=0~,
\ee
where $\hat{\pi}_\alpha=-i \frac{\partial}{\partial\alpha}$ and
$\hat{\pi}_\phi=-i \frac{\partial}{\partial\phi}$, or
\be\label{wdwalphi}
\left\{\frac{1}{2}\partial^2_{\alpha}-\frac{1}{2}\partial^2_{\phi}+V_0 e^{6\alpha-3\sqrt{2}\phi}   \right\}\Psi(\alpha,\phi)=0~.
\ee

In the de Broglie-Bohm (dBB) quantum theory, the universe is described by an objective actual configuration $(\alpha(t),\phi(t))$ whose motion is guided by a wave-function $\Psi$ satisfying the WDW equation \eqref{wdwalphi}.
The configuration $(\alpha(t),\phi(t))$ evolves according to the guidance equations
\begin{align}\label{guidance_ap}
&\dot{\alpha}=\frac{1}{m_\alpha}\mathfrak{Im}\left(\frac{\partial_\alpha\Psi}{\Psi}\right)=\frac{1}{m_\alpha}\partial_\alpha S(\alpha,\phi)\,,&\nonumber\\
&\dot{\phi}=\frac{1}{m_\phi}\mathfrak{Im}\left(\frac{\partial_\phi\Psi}{\Psi}\right)=\frac{1}{m_\phi}\partial_\phi S(\alpha,\phi)\,,&
\end{align}
where $\mathfrak{Im}$ denotes the imaginary part, $\Psi=R e^{iS}$, $m_\alpha=-e^{3\alpha}$ and $m_\phi=e^{3\alpha}$.

The problem therefore amounts to finding a physically suitable solution of Eq.~\eqref{wdwalphi} yielding bouncing trajectories satisfying (\ref{guidance_ap}).
First, we need to obtain a basis of solutions of the WDW equation, and for that we need to introduce other coordinate systems.
\subsection{First coordinate system}
A first possibility is the coordinate system used in \cite{dks}
\begin{align}\label{uvsys}
u=&2\frac{\sqrt{2V_0}}{3}e^{\tilde{X}}(\cosh{X}+\frac{1}{\sqrt{2}}\sinh{X})\,,&\nonumber\\
v=&2\frac{\sqrt{2V_0}}{3}e^{\tilde{X}}(\sinh{X}+\frac{1}{\sqrt{2}}\cosh{X})\,,&
\end{align}
where $X=3(\phi-\frac{1}{\sqrt{2}}\alpha)$ and  $\tilde{X}=3(\alpha-\frac{1}{\sqrt{2}}\phi)$.
A similar coordinate system proved itself useful in a study of the Big Rip in the dBB theory \cite{npndmo}.

It should be noted that the domain of $(u,v)$ is restricted to $u\geq 0$ and $u^2-v^2\geq 0$.
In the $(u,v)$ coordinate system, the WDW equation becomes
\be
\left(\partial^2_{u}-\partial^2_{v}+1\right)\Psi(u,v)=0~,
\ee
with $\Psi(u,v)$ defined on the above domain.
It is therefore a Klein-Gordon equation, whose basis solutions are known:
\be
\varphi_{[E,k]}(u,v)=e^{i{(k\,v-E\,u)}}\,,
\ee
with $E=\pm\sqrt{k^2+1}$ and $k$ real.

We will refer to these solutions as KG-type solutions (where KG stands for Klein-Gordon).
\subsection{Second coordinate system}
Another coordinate transformation is the following one:
\be\label{newcvar}
x_1=\biggl(\alpha-\frac{\phi}{\sqrt{2}}\biggr)2\sqrt{V_0}\quad\quad
x_2=\biggl(\phi-\frac{\alpha}{\sqrt{2}}\biggr)2\sqrt{V_0}~\,,
\ee
or
\be
\alpha=\frac{1}{\sqrt{V_0}}\biggl(x_1+\frac{x_2}{\sqrt{2}}\biggr)\quad\quad
\phi=\frac{1}{\sqrt{V_0}}\biggr(x_2+\frac{x_1}{\sqrt{2}}\biggl)~.
\ee

In this new coordinate system, the classical dust attractor $\alpha=\sqrt{2} \phi +$const. is at $x_2=0$, the stiff matter behavior $\alpha=\pm\phi+$const. is mapped to $x_1=\pm x_2+$ const., respectively, and the $w=-1$ dark energy transition point is mapped to $x_1=-\sqrt{2} x_2 + $const..

The WDW equation becomes
\be
\left\{\partial^2_{x_1}-\partial^2_{x_2}+e^{\frac{3  x_1}{\sqrt{V_0}}}\right\}\Psi(x_1,x_2)=0~.
\ee
We can solve this equation by the method of separation of variables. We write
\be
\Psi(x_1,x_2)=f_{1k}(x_1)f_{2k}(x_2)\quad\textrm{with}\quad f_{2k}(x_2)=e^{ikx_2}~.
\ee
Then the WDW equation leads to the equation
\be\label{wdwpot}
\left\{\partial^2_{x_1}+k^2+e^{{2\gamma x_1}}\right\}f_{1k}(x_1)=0~,
\ee
where we have introduced a parameter $\gamma=3 /(2\sqrt{V_0})$.
The solution to that equation is in \cite{cofanpn}.
In the present case, it is
$f_{1k}(x_1)=J_{\pm i\frac{k}{\gamma}}(\frac{e^{\gamma x_1}}{\gamma})$ or
$f_{1k}(x_1)=Y_{\pm i\frac{k}{\gamma}}(\frac{e^{\gamma x_1}}{\gamma})$, where $J$ and $Y$ are Bessel functions of real and purely imaginary order.

Overall, the basis functions for the WDW equation are of the form
\be
Z_{\pm i\frac{k}{\gamma}}\biggl(\frac{e^{\gamma x_1}}{\gamma}\biggr)e^{ik x_2}=
Z_{\pm i\nu}(\mathcal{E})e^{ik x_2}\,,
\ee
where $Z=J$ or $Z=Y$. We will refer to these solutions as Bessel-type solutions.

Useful relations regarding the Bessel functions of imaginary order can be found in the papers by Dunster \cite{dunster} and
Chapman \cite{chapman}.
To introduce these relations, the Bessel functions are first decomposed into their real and imaginary parts, for example $J_{i\nu}=J_{i\nu,r}+i J_{i\nu,i}$.
The useful identities are then
\begin{align}
&Y_{i\nu,r}=\coth\biggl(\frac{\pi\nu}{2}\biggr)J_{i\nu,i}\,,&
&Y_{i\nu,i}=-\tanh\biggl(\frac{\pi\nu}{2}\biggr)J_{i\nu,r}\,,&
\end{align}
and
\begin{align}
&J_{i\nu,r}=\cosh\biggl(\frac{\pi\nu}{2}\biggr)\tilde{J}_{\nu}\,,&
&J_{i\nu,i}=\sinh\biggl(\frac{\pi\nu}{2}\biggr)\tilde{Y}_{\nu}\,,&
\end{align}
where $\tilde{J}$ and $\tilde{Y}$ are real functions (defined in \cite{dunster} as $F$ and $G$),
invariant under $\nu\rightarrow -\nu$.
Thanks to these relations, we can write the Bessel functions of imaginary order, $J$ and $Y$,  as
\begin{align}
J_{i\nu}(\mathcal{E})=&\cosh\biggl(\frac{\pi\nu}{2}\biggr)\tilde{J}_{\nu}(\mathcal{E})+i\sinh\biggl(\frac{\pi\nu}{2}\biggr)\tilde{Y}_{\nu}(\mathcal{E})\,,&\\
Y_{i\nu}(\mathcal{E})=&\cosh\biggl(\frac{\pi\nu}{2}\biggr)\tilde{Y}_{\nu}(\mathcal{E})-i\sinh\biggl(\frac{\pi\nu}{2}\biggr)\tilde{J}_{\nu}(\mathcal{E})~.&
\end{align}
Conversely, we have that
\begin{align}
\tilde{J}_\nu(\mathcal{E})=\sech\biggl(\frac{\pi\nu}{2}\biggr)\mathfrak{Re}(J_{i\nu}(\mathcal{E}))\,,\nonumber\\
\tilde{Y}_\nu(\mathcal{E})=\sech\biggl(\frac{\pi\nu}{2}\biggr)\mathfrak{Re}(Y_{i\nu}(\mathcal{E}))~.
\end{align}

We can see that $\{\tilde{J}_{\frac{|k|}{\gamma}}(\mathcal{E})e^{i k x_2},\tilde{Y}_{\frac{|k|}{\gamma}}(\mathcal{E})e^{i k x_2}\}$ is another possible choice of basis for solutions of the WDW equation.

For $\mathcal{E}\gg1$, we have that
\be\label{besapprox1}
\tilde{J}_{\nu}(\mathcal{E})\approx\sqrt{\frac{2}{\pi \mathcal{E}}}\cos\biggl(\mathcal{E}-\frac{\pi}{4}\biggr)\,,\quad
\tilde{Y}_{\nu}(\mathcal{E})\approx\sqrt{\frac{2}{\pi \mathcal{E}}}\sin\biggl(\mathcal{E}-\frac{\pi}{4}\biggr)\,.
\ee
These approximations render the equations simple, and they do not depend on $\nu$.
Furthermore, we can assume that
\begin{align}\label{besapprox2}
&(\tilde{J}_{\nu}(\mathcal{E}))'\approx
\gamma\sqrt{\frac{2}{\pi}}\left[-\sqrt{\mathcal{E}}\sin\biggl(\mathcal{E}-\frac{\pi}{4}\biggr)-\frac{1}{2\sqrt{\mathcal{E}}}\cos\biggl(\mathcal{E}-\frac{\pi}{4}\biggr)\right]\,,&\nonumber\\
&(\tilde{Y}_{\nu}(\mathcal{E}))'\approx
\gamma\sqrt{\frac{2}{\pi}}\left[\sqrt{\mathcal{E}}\cos\biggl(\mathcal{E}-\frac{\pi}{4}\biggr)-\frac{1}{2\sqrt{\mathcal{E}}}\sin\biggl(\mathcal{E}-\frac{\pi}{4}\biggr)\right]~,&
\end{align}
where a prime means $\partial /\partial x_1$.

For $\mathcal{E}\ll1$ and $\nu>0$, we have that
\begin{align}
\tilde{J}_{\nu}(\mathcal{E})=&\left[\frac{2\tanh(\frac{\pi\nu}{2})}{\pi\nu}\right]^{\frac{1}{2}}\cos\left[\nu\ln\biggl(\frac{\mathcal{E}}{2}\biggr)-\arg \Gamma(1+i\nu)\right]~,&\label{besapprox3}\\
\tilde{Y}_{\nu}(\mathcal{E})=&\left[\frac{2\coth(\frac{\pi\nu}{2})}{\pi\nu}\right]^{\frac{1}{2}}\sin\left[\nu\ln\biggl(\frac{\mathcal{E}}{2}\biggr)-\arg \Gamma(1+i\nu)\right]~.\label{besapprox4}&
\end{align}
Only in the limit $\nu\rightarrow 0$ will the above approximation become simpler; we start from (\ref{besapprox3}) and (\ref{besapprox4}), which are valid for
$\nu>0$ ($\nu=|k|/\gamma$). In the limit where $\nu$ is small, we have that
$\tanh(\frac{\pi\nu}{2})\approx\frac{\pi\nu}{2}$ and $\arg (\Gamma(1+i\nu))=-\mathfrak{e}\nu$ (where $\mathfrak{e}=0.57\ldots$ is Euler's constant). Therefore
\begin{align*}
\tilde{J}_{\nu}(\mathcal{E})\approx \cos\biggl[\nu\log\biggl(\frac{\mathcal{E}}{2}\biggr)+\mathfrak{e}\nu\biggr]=\cos\biggl[\nu\log\biggl(\frac{e^{\gamma x_1}}{2\gamma}\biggr)+\mathfrak{e}\nu\biggr]=\nonumber\\
\cos\biggl[|k|x_1-\frac{|k|}{\gamma}(\log(2\gamma)-\mathfrak{e})\biggr]
=\cos\biggl[k x_1-\frac{k}{\gamma}(\log(2\gamma)-\mathfrak{e})\biggr]\,.\end{align*}
We denote $(\log(2\gamma)-\mathfrak{e})/\gamma$ by $C_\gamma$. Similarly we find that
\begin{align}
\tilde{Y}_{\nu}(\mathcal{E})\approx \frac{2}{\pi\nu}\sin\biggl[|k|x_1-\frac{|k|}{\gamma}(\log(2\gamma)-\mathfrak{e})\biggr]\nonumber\\
=\frac{2\gamma}{\pi k}\sin(k x_1-k C_\gamma)\,.\end{align}
For example, we have that
\be
e^{i k C_\gamma}\biggl[\tilde{J}_{\nu}(\mathcal{E})+i\frac{\pi k}{2\gamma}\tilde{Y}_{\nu}(\mathcal{E})\biggr]\approx e^{i  k x_1}\,,
\ee
and
\be
e^{-i k C_\gamma}\biggl[\tilde{J}_{\nu}(\mathcal{E})-i\frac{\pi k}{2\gamma}\tilde{Y}_{\nu}(\mathcal{E})\biggr]\approx e^{-i k x_1}\,.\ee

In the appendix we give a relation connecting the $J_{0}(\mathcal{E})$ and $Y_{0}(\mathcal{E})$ solutions to the KG-type solutions.

\section{The quantum bouncing trajectories}
\subsection{The dBB guidance equations}
We will use the $(x_1,x_2)$ coordinate system to plot the trajectories.
Indeed $x_1$ and $x_2$ appear naturally for KG-type and Bessel-type solutions (for KG-type solutions, one notes that $X=\gamma x_2$ and $\tilde{X}=\gamma x_1$ where
$X$ and $\tilde{X}$ are found in (\ref{uvsys})).
At this stage we do not know which basis of solutions will be more useful to obtain bouncing trajectories.
We will explore the space of wave-functions of KG and Bessel type and their associated dBB trajectories.

The solutions of the WDW equation of interest (that is, which exhibit bouncing behavior) will be denoted by
\be
\Psi_{k}=\mathcal{R}_k e^{i\mathfrak{S}}_k \,.
\ee
The associated dBB velocity flow is
\be\label{vitesse}
v_{k,1}=\frac{1}{m_1}\partial_1\mathfrak{S}_k \,,v_{k,2}=\frac{1}{m_2}\partial_2\mathfrak{S}_k \,,
\ee
where $m_1=-  e^{3\alpha}/{(2V_0)}$ and $m_2=-m_1$.
We also define the normed dBB velocity flow as
\be\label{nvf}
\tilde{v}_1(x_1,x_2)=\frac{v_1}{v_1^2+v^2_2}~,\quad
\tilde{v}_2(x_1,x_2)=\frac{v_2}{v_1^2+v^2_2}~.
\ee
\subsection{The classical trajectories}
The KG-type solutions are
\be
\varphi_{[E,k]}(u,v)=e^{i(k v - E u)}
\ee
where $E=\pm \sqrt{k^2+1}$ and $u$ and $v$, as functions of $x_1$ and $x_2$, are given by
\begin{align}
u=&\frac{e^{\gamma x_1} }{\gamma}[\sinh{(\gamma x_2)}+\sqrt{2}\cosh{(\gamma x_2)}]\,,&\label{ux1x2}\\
v=&\frac{e^{\gamma x_1} }{\gamma}[\cosh{(\gamma x_2)}+\sqrt{2}\sinh{(\gamma x_2)}]\,.&\label{vx1x2}\end{align}
We also have that
\begin{align}
\frac{\partial_1\varphi}{\varphi}=&i e^{\gamma x_1}[\cosh{(\gamma x_2)}(\sqrt{2}k-E)+\sinh{(\gamma x_2)}(k-\sqrt{2}E)]\,,&\nonumber\\
\frac{\partial_2\varphi}{\varphi}=&i e^{\gamma x_1}[\sinh{(\gamma x_2)}(\sqrt{2}k-E)+\cosh{(\gamma x_2)}(k-\sqrt{2}E)]\,.&
\end{align}
Given the last relations, it is easy to obtain the dBB trajectories for a superposition of KG-type solutions
\be
\Psi=\sum_{n} c_n\varphi_n=\sum_{n} c_n \varphi_{[E_n,k_n]}(u,v)\,.
\ee
One has that
\begin{eqnarray}\label{kgdbb}
v_1=\frac{1}{m_1}
\mathfrak{Im}\left(
\frac{\sum_{n}c_n(\varphi_n^{-1}\partial_1\varphi_n)\varphi_n}{\Psi}
\right)\,,
\nonumber\\
v_2=\frac{1}{m_2}
\mathfrak{Im}\left(
\frac{\sum_{n}c_n(\varphi_n^{-1}\partial_2\varphi_n)\varphi_n}{\Psi}
\right)\,,
\end{eqnarray}
where $m_2= e^{3\alpha}/(2 V_0)$ and $m_1=-m_2$.

\begin{figure}[H]
\begin{center}
\includegraphics[width=0.45\textwidth]{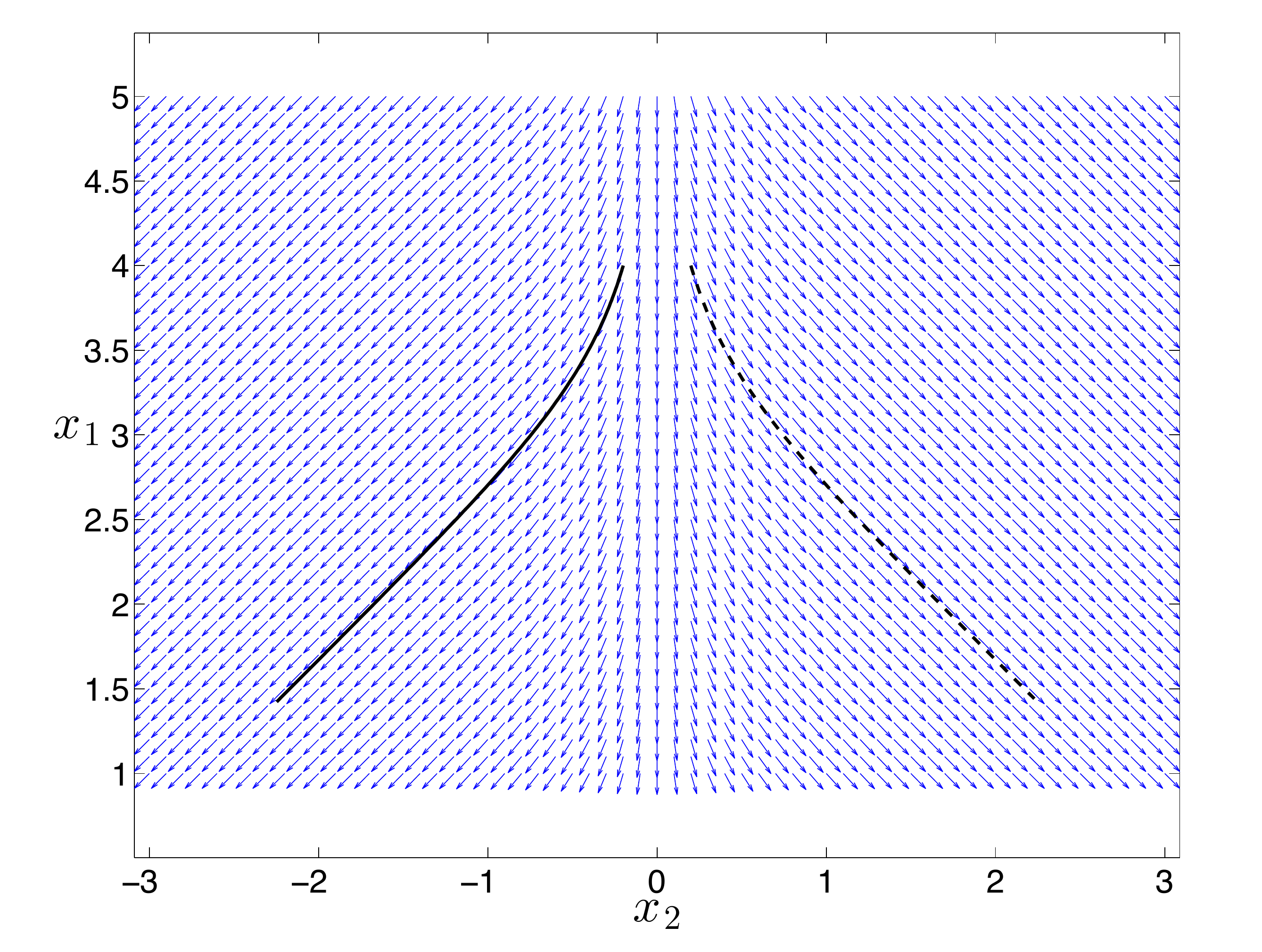}\\
\includegraphics[width=0.45\textwidth]{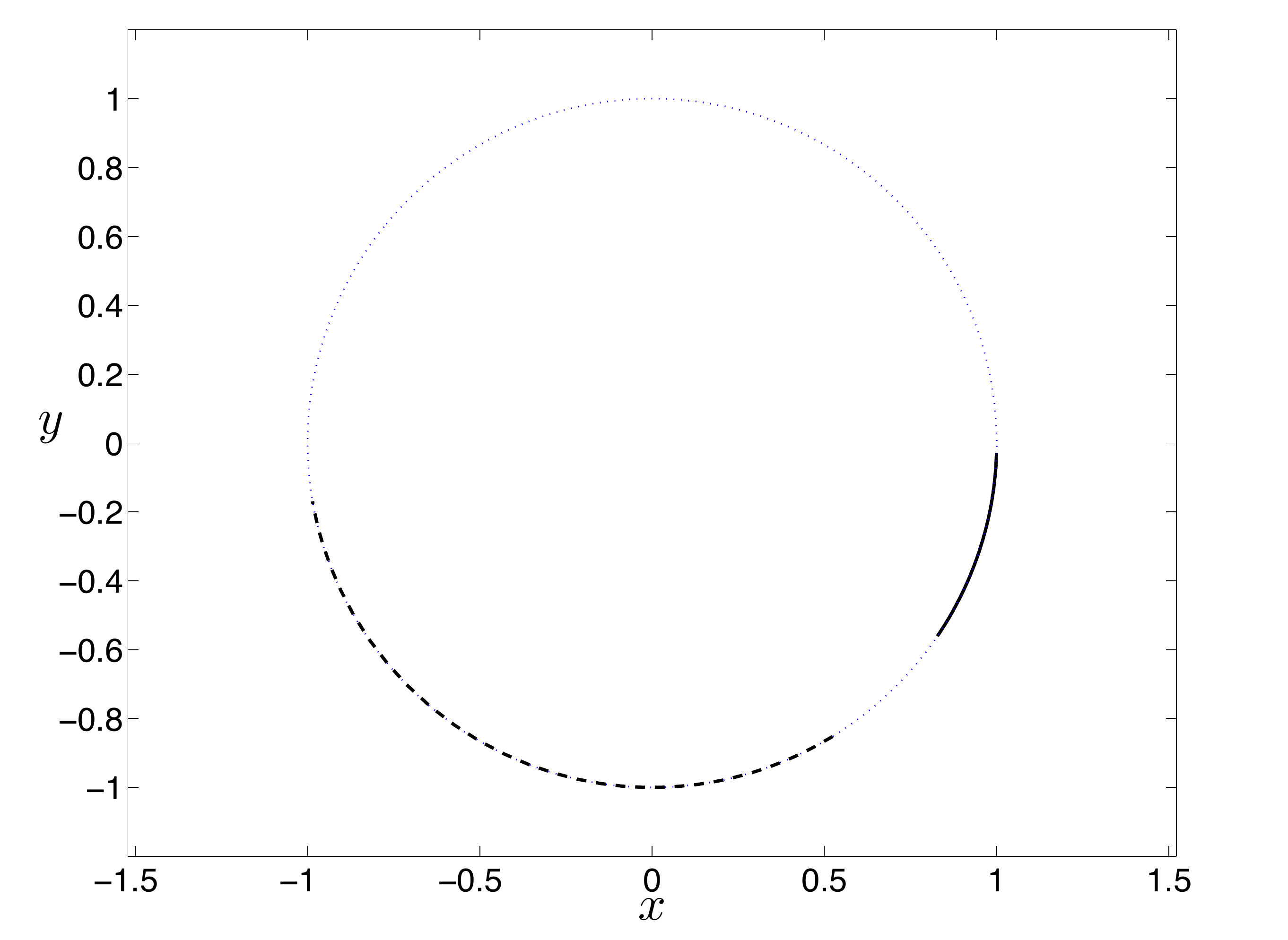}
\end{center}
\caption{Top-part of the figure: contracting dBB trajectories, obtained by solving (\ref{kgdbb})
for the WDW solution $e^{i(-v + \sqrt{2} u)}$, together with the normed dBB velocity flow (see (\ref{nvf})).
One trajectory (dashed line) starts at $(4,0.2)$ and runs from $t=0$ to $t=310$,
the other one (continuous line) starts at $(4,-0.2)$ and runs from $t=0$ to $t=45.7315$.
Bottom-part of the figure: representation of the 2 trajectories in the phase-space of Heard and Wands ($x$ and $y$ coordinates defined at (\ref{hewaxy})).}
\label{fig3}
\end{figure}

\begin{figure}[H]
\begin{center}
\includegraphics[width=0.45\textwidth]{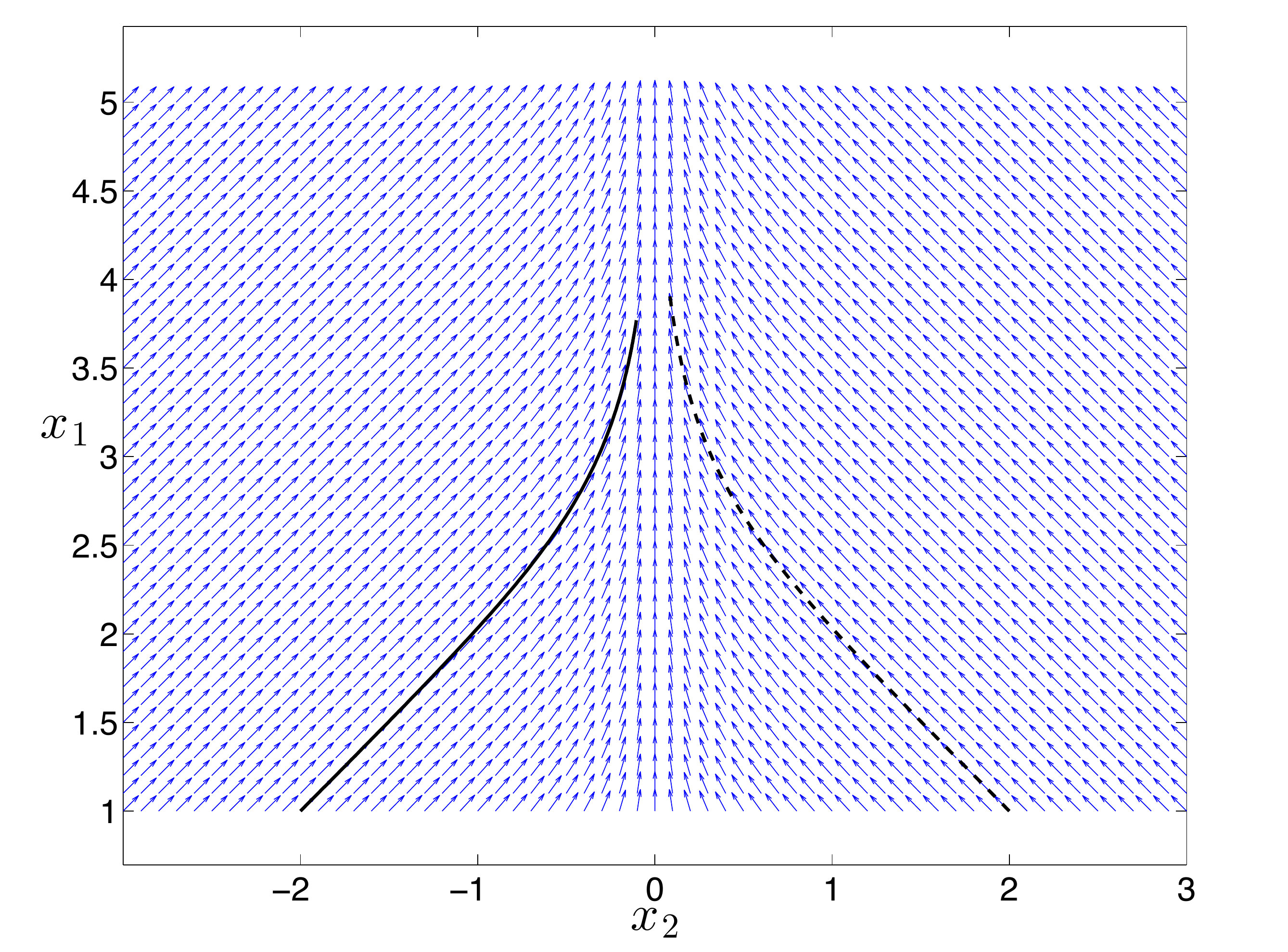}\\
\includegraphics[width=0.45\textwidth]{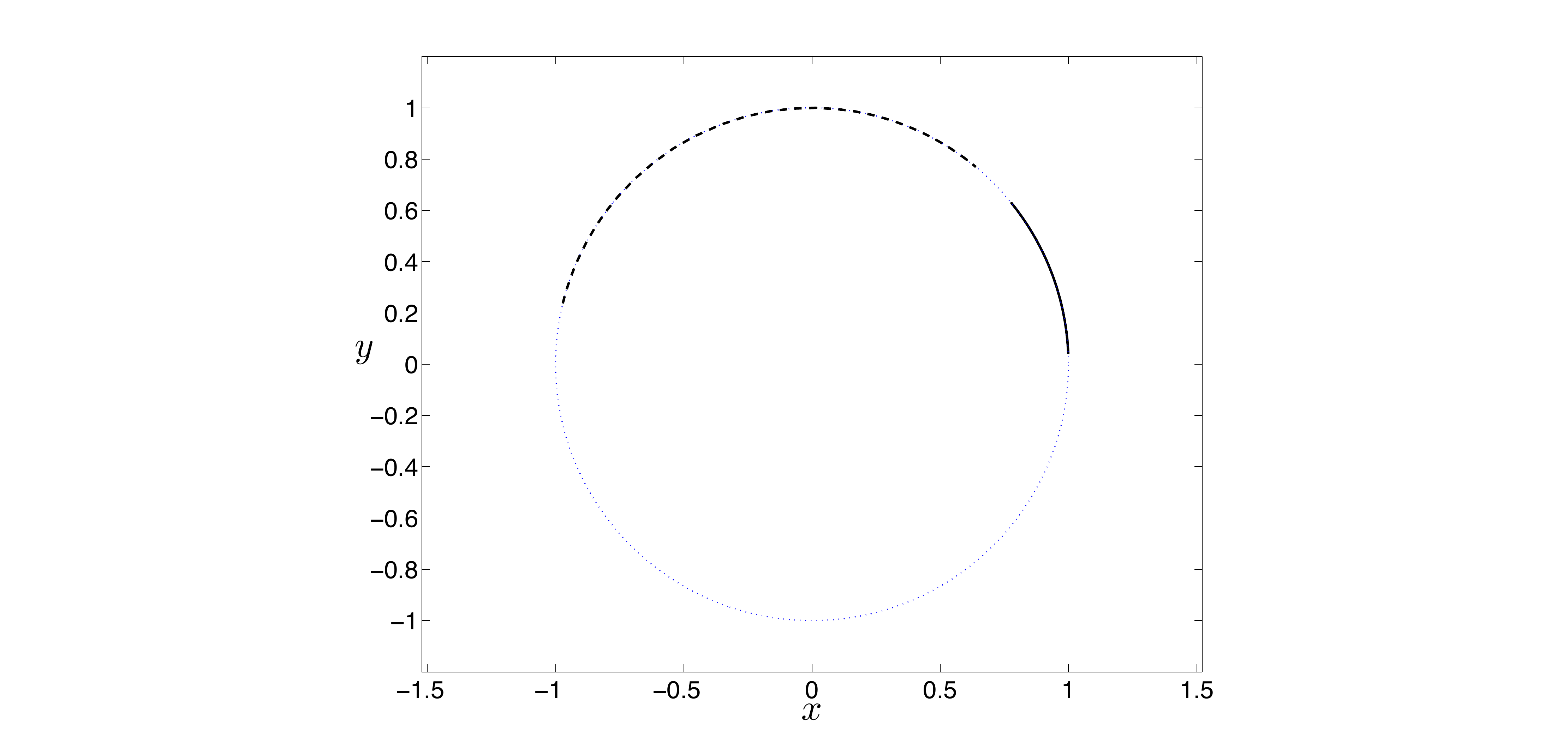}
\end{center}
\caption{Top-part of the figure: expanding dBB trajectories, obtained by solving (\ref{kgdbb})
for the WDW solution $e^{i(v - \sqrt{2} u)}$, together with the normed dBB velocity flow (see (\ref{nvf})).
One trajectory (dashed line) starts at $(1,2)$ and runs from $t=0$ to $t=200$,
the other one (continuous line) starts at $(1,-2)$ and runs from $t=0$ to $t=50$.
Bottom-part of the figure: representation of the 2 trajectories in the phase-space of Heard and Wands ($x$ and $y$ coordinates defined at (\ref{hewaxy})).}
\label{fig4}
\end{figure}

\begin{figure}[H]
\begin{center}
\includegraphics[width=0.45\textwidth]{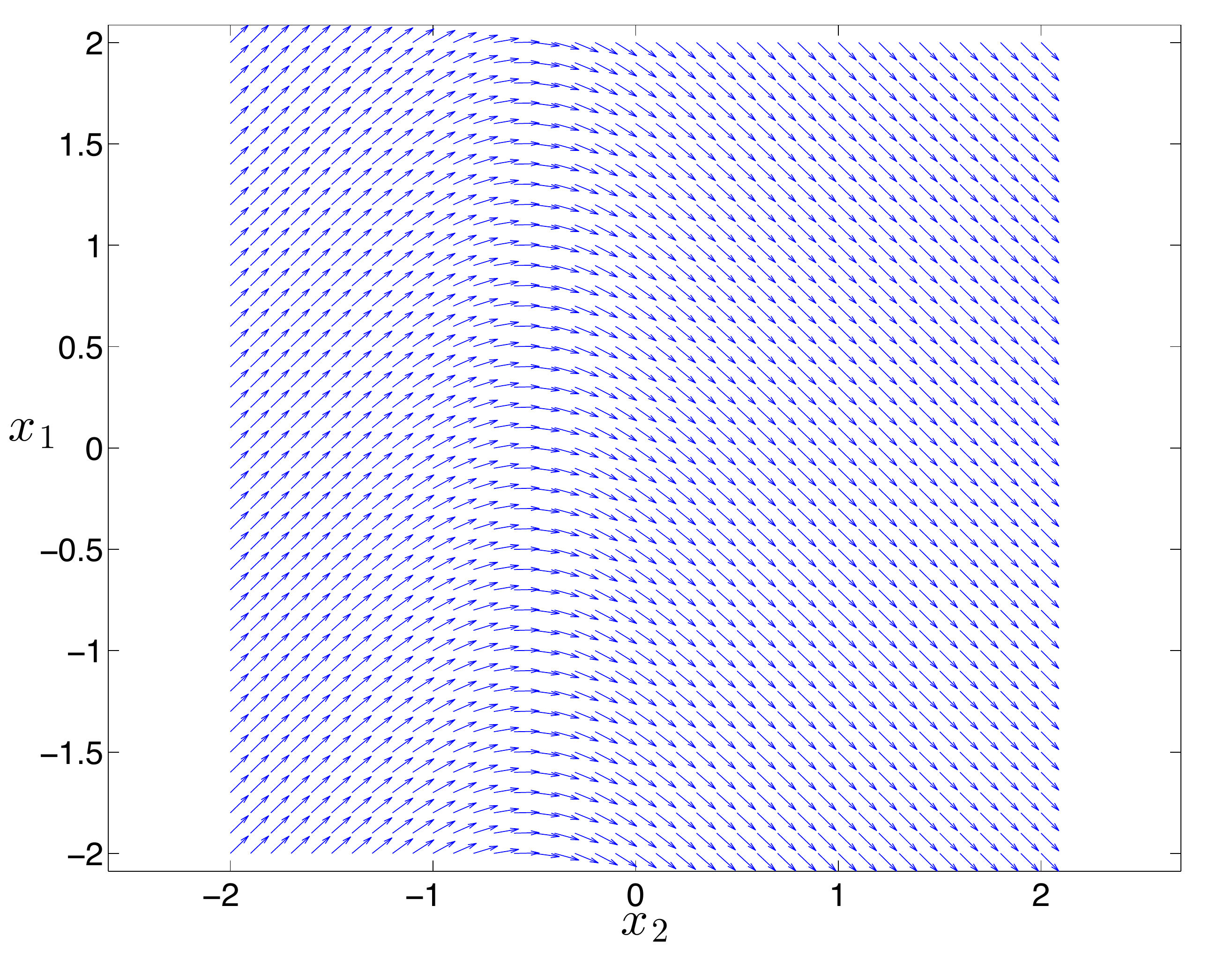}
\end{center}
\caption{Normed dBB velocity flow for $e^{i(v - \sqrt{2} u)}+e^{i(-v + \sqrt{2} u)}$.}
\label{fig5}
\end{figure}

For example, taking a single plane wave solution with $k=-1$ and $E=-\sqrt{2}$, we obtain classical contracting universes, which start in the vicinity of the dust attractor and move towards the singularity, passing or not through the dark energy phase.
The trajectories of two such universes are plotted in Fig. (\ref{fig3}), together with the normed dBB velocity flow.
On the other hand, for $k=1$ and $E=\sqrt{2}$, we have expanding universes; the trajectories and the normed dBB flow are plotted in Fig. (\ref{fig4}).
If we superpose these two KG-type solutions, with positive and negative energies,
we do not get a bounce, but an expanding universe, which reaches a maximal radius and undergoes a contraction (see Fig. (\ref{fig5})).
That is the most interesting scenario coming from the superposition of two KG-type solutions - no bounce seems to arise by superposing two KG-type solutions.
\subsection{The semi-classical approximation}
In order to get bounces, we will have to look into more involved solutions, like Gaussian packets.
\subsubsection{A single packet}
We consider a Gaussian packet of KG-type solutions centered around $k$:
\begin{eqnarray}\label{narrow_gau}
\Psi^\sigma_{[E,k]}(u,v):=R e^{iS}=\nonumber\\\frac{1}{\sqrt{\sqrt{\pi}\sigma }}\int dq \exp\left[-\frac{(q-k)^2}{2\sigma^2}\right] e^{i(q v - E_q u)}\,.
\end{eqnarray}
As far as we know, there is no analytical expression for such a Gaussian packet, so we take a semi-classical (narrow) packet approximation.
In that approximation, we have that \cite{dks,npndmo}
\begin{align}\label{gaussianapprox}
R=&\frac{1}{(1+\sigma^4{S''_0}^2)^{1/4}}\exp\biggl[-\frac{{S'_0}^2\sigma^{-2}}{2(\sigma^{-4}+{S''_0}^2)}\biggr]\, ,&\\
S=&S_0-\frac{{S'_0}^2{S''_0}}{2(\sigma^{-4}+{S''_0}^2)}+\frac{1}{2}\arctan(\sigma^2  S''_0)~,&
\end{align}
where we are assuming that $\sigma/k \ll 1$.
In the present case, $S_0=kv-Eu$ (where $E$ can be positive or negative, and $E'=k/E$) and we have that $S'_0=\displaystyle\frac{vE-ku}{E}$ and $S''_0=-\displaystyle\frac{u}{E^3}$.

Therefore the approximated action becomes
\be\label{s_approx}
S=(kv-Eu)+\frac{\sigma^4(vE-ku)^2uE}{2(E^6+\sigma^4 u^2)}+\frac{1}{2}\arctan\left(-\frac{u\sigma^2}{E^3}\right).
\ee
The gradients of the phase are:
\begin{align}
\partial_u S=-E-\frac{E^3\sigma^2}{2(E^6+\sigma^4 u^2)}+\nonumber\\\frac{E\sigma^4(Ev-ku)(E^7v-3E^6ku-E\sigma^4u^2v-k\sigma^4u^3)}{2(E^6+\sigma^4 u^2)^2}\nonumber\\
\partial_v S=k+\frac{E^2\sigma^4u(Ev-ku)}{E^6+\sigma^4 u^2} \, .
\end{align}
\subsubsection{A superposition of two Gaussian packets}
We have that $R=\frac{1}{A^{\frac{1}{4}}}\exp(-\frac{{S'_0}^2}{2\sigma^{-2}A})$ with $A=1+\sigma^4 {S''_0}^2$.
We introduce $B$ and $C$ through the relation
\be{S'_0}^2=(\frac{vE-ku}{E})^2=(v^2+\frac{k^2u^2}{E^2})-2uv\frac{k}{E}~=B-C.\ee
We see that $A$ and $B$ are invariant under $E\rightarrow -E$ or $k\rightarrow -k$.
But $C=2uv\displaystyle\frac{k}{E}$ is only invariant under both operations combined, and changes sign under a single operation.
We introduce the positive function $F=\bigg|\frac{k}{E}\bigg|\frac{1}{\sigma^{-2}A}$.

We will have to distinguish several cases when we superpose 2 Gaussian packets.
From now on, and until the end of this subsection, it is assumed that $E$ and $k$ are positive.
The following terminology is used to differentiate an individual packet:
\begin{itemize}
\item Type 1: Positive-energy, centered at $k$ \\ ($S_1=S(E,k)$ where S is defined at (\ref{s_approx})):
\begin{equation}
\Phi_1=A^{-\frac{1}{4}}\exp\left(-{\frac{B}{2 \sigma^{-2} A}}\right)\exp\left(F u v\right)e^{iS_1}\,.
\end{equation}
\item Type 2: Negative-energy, centered at $k$ \\ ($S_2=S(-E,k)$):
\be
\Phi_2=A^{-\frac{1}{4}}\exp\left(-{\frac{B}{2 \sigma^{-2} A}}\right)\exp\left(-F u v\right)e^{iS_2}\,.\ee
\item Type 3: Positive-energy, centered at $-k$ \\ ($S_3=S(E,-k)$):
\be
\Phi_3=A^{-\frac{1}{4}}\exp\left(-{\frac{B}{2 \sigma^{-2} A}}\right)\exp\left(-F u v\right)e^{iS_3}\,.\ee
\item Type 4: Negative-energy, centered at $-k$ \\ ($S_4=S(-E,-k)$):
\be
\Phi_4=A^{-\frac{1}{4}}\exp\left(-{\frac{B}{2 \sigma^{-2} A}}\right)\exp\left(F u v\right)e^{iS_4}\,.\ee
\end{itemize}
\paragraph{Superposition $\Phi_1+c\Phi_2$ with $k$ positive.}
The wave-function has the structure
\begin{align}\label{bigpsi1}
\Psi_1=A^{-\frac{1}{4}}\exp\left(-{\frac{B}{2 \sigma^{-2} A}}\right)\times\nonumber\\
\left\{\exp(F u v)e^{iS_1}+c\exp(-F u v)e^{iS_2}\right\}=\nonumber\\
(\ldots)\left\{\exp(F u v)e^{iS_1}+c\exp(-F u v)e^{iS_2}\right\}\,,
\end{align}
where $(\ldots)$, from now on in this section,
will denote a real expression of no relevance for the guidance equation.

We have to make the distinction between $v$ positive and negative (we do this distinction because there will be an exponential involving $F u v$, $F$ and $u$ being positive, in order to avoid divergences).\\
\underline{Case I: $v<0$}. We write $\Psi_1$ as
\begin{align}
(\ldots)e^{iS_2}\left\{\frac{1}{c}\exp(2 F u v)e^{i(S_1-S_2)}+1\right\}=\nonumber\\
(\ldots)e^{iS_2}\left\{b\exp(G)e^{i(S_1-S_2)}+1\right\}~,
\end{align}
where $b=c^{-1}$ and $G=2 F u v$.
$\Psi_1$ can be further written as
\be
(\ldots)e^{i\biggl[S_2+\arctan\biggl(\frac{f}{g}\biggr)+K\biggr]}~,
\ee
where
\begin{align}
f=b\exp(G)\sin(S_1-S_2)\,,\nonumber\\
g=1+b\exp(G)\cos(S_1-S_2)\, ,
\end{align}
and $K$ is non-zero if $g<0$ ($\pm\pi$ depending on the sign of $f$).
The gradient of the total phase is given by
\begin{align}\label{grads_2gau}
\mathfrak{S}'_1=S'_2+\frac{b \exp(G)}{f^2+g^2}\big[G'\sin(S_1-S_2)+\nonumber\\\cos(S_1-S_2)(S_1-S_2)'+b\exp(G)(S_1-S_2)'\big]~,
\end{align}
where $f^2+g^2=1+2b\exp(G)\cos(S_1-S_2)+b^2\exp(2G)$ and where the prime indicates a derivative with respect to $u$ or $v$.

The phase difference is
\be
S_1-S_2=-2 E u +\frac{E\sigma^4 u (E^2 v^2 + k^2 u^2)}{E^6+\sigma^4 u^2}-\arctan\left(\frac{\sigma^2 u}{E}\right)
\ee
and the gradient is given by
\begin{align}
\partial_u(S_1-S_2)=-2 E-\frac{E^3\sigma^2}{E^6+\sigma^4u^2}+\nonumber\\\frac{E\sigma^4(E^8v^2+3E^6k^2u^2-E^2\sigma^4u^2v^2+k^2\sigma^4u^4)}{(E^6+\sigma^4u^2)^2}~,\nonumber\\
\partial_v(S_1-S_2)=2\frac{E^3\sigma^4 u v}{E^6+\sigma^4 u^2}~.
\end{align}

The gradient of $G$ is
\be
\partial_u G=2 E^5k\sigma^2v\frac{(E^6-\sigma^4u^2)}{(E^6+\sigma^4u^2)^2}\,,\quad
\partial_v G=2\frac{E^5k\sigma^2u}{E^6+\sigma^4u^2}~.
\ee
\underline{Case II: $v>0$}. We write $\Psi_1$ as
\be
(\ldots)e^{iS_1}\left\{1+c\exp(-2 F u v)e^{i(S_2-S_1)}\right\}~.
\ee
The rest is similar to the previous case (\ref{grads_2gau}), except that we interchange 1 and 2, substitute $b$ by $c$, and $G$ by $-G$. After some rewriting, this gives
\begin{align}\label{grads_2gau2}
\mathfrak{S}'_1=S'_1+\frac{c \exp(-G)}{(f^2+g^2)_{G\rightarrow -G}}
\big[G'\sin(S_1-S_2)\nonumber\\-\cos(S_1-S_2)(S_1-S_2)'-c\exp(-G)(S_1-S_2)'\big]~.
\end{align}

In the figures (\ref{fig6}) and (\ref{fig7}) below, we present the Bohmian trajectories corresponding to the wave solution $\Psi_1$.

\begin{figure}[H]
\begin{center}
\includegraphics[width=0.45\textwidth]{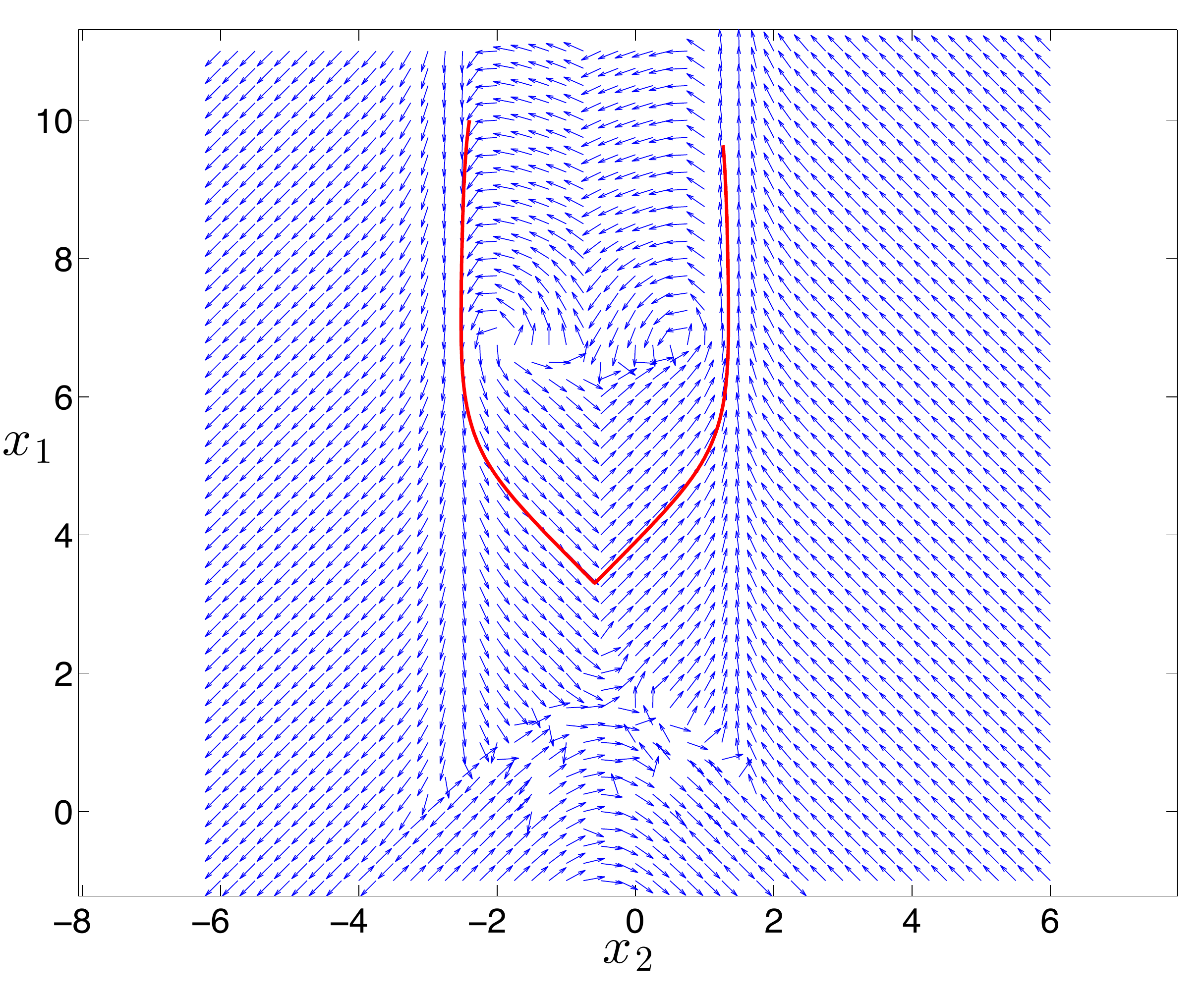}
\end{center}
\caption{Normed dBB flow for $\Psi_1$ (see (\ref{bigpsi1}) with $\sigma=0.1$, $k=10$ and $E=\sqrt{101}$), together with the dBB trajectory,
obtained by solving (\ref{vitesse}), thanks to the relations (\ref{grads_2gau}) and (\ref{grads_2gau2}), for the initial condition $(10,-2.4)$ and $t\in[0,10^{7}]$.}
\label{fig6}
\end{figure}
\begin{figure}[H]
\begin{center}
\includegraphics[width=0.45\textwidth]{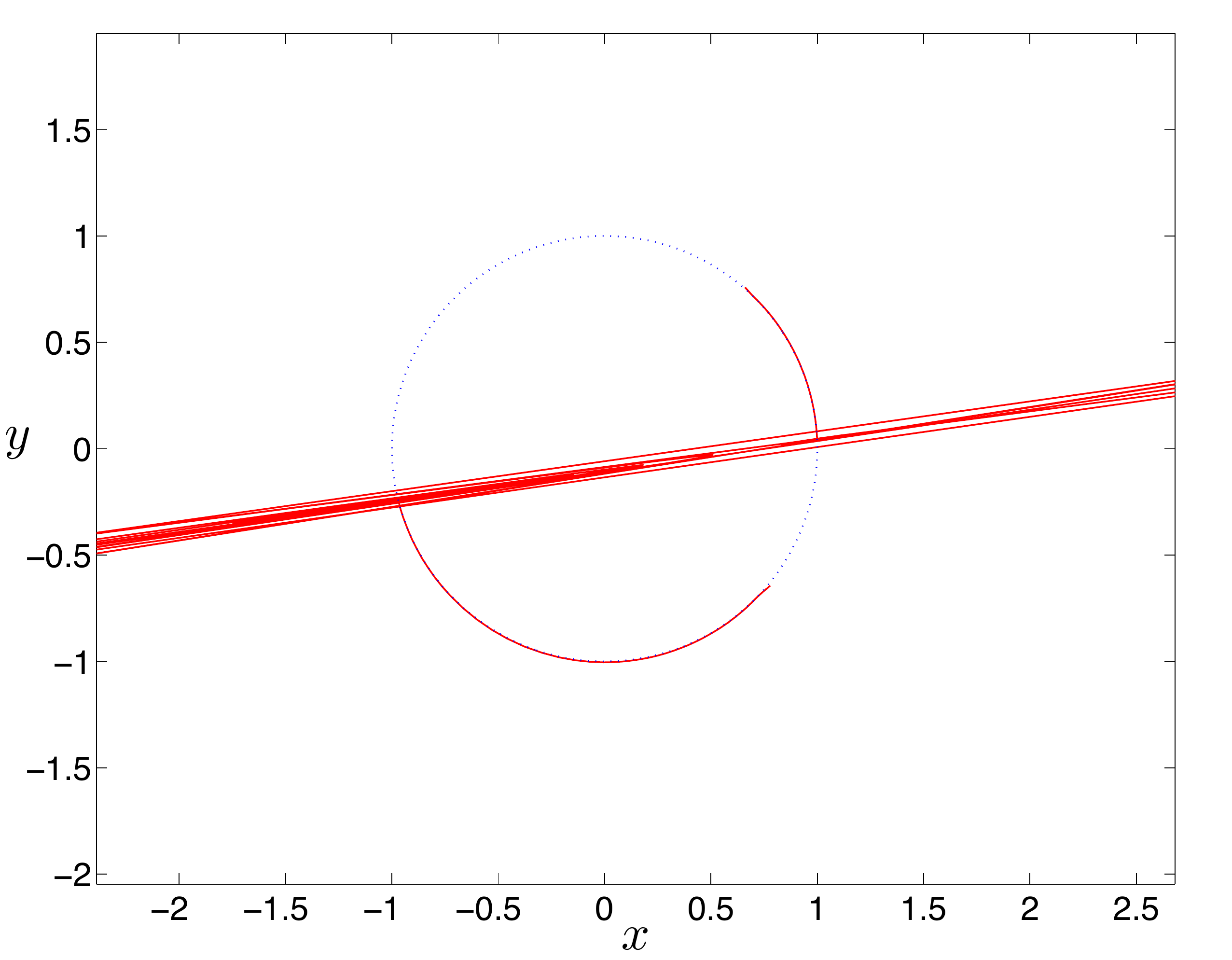}
\end{center}
\caption{Trajectory of Fig. (\ref{fig6}) visualized in the phase-space of Heard and Wands. The universe starts in the neighborhood
of $(\frac{1}{\sqrt{2}},-\frac{1}{\sqrt{2}})$ and ends ups in the neighborhood of $(\frac{1}{\sqrt{2}},\frac{1}{\sqrt{2}})$.
We have that $x\in[-42,158]$ and $y\in[-6,21]$.}
\label{fig7}
\end{figure}
\paragraph{Superposition $\Psi_2=\Phi_4+c\Phi_3$.}
We have that
\begin{align}\label{bigpsi2}
\Psi_2=A^{-\frac{1}{4}}\exp\left(-{\frac{B}{2 \sigma^{-2} A}}\right)\times\nonumber\\
\left\{\exp(F u v)e^{iS_4}+c\exp(-F u v)e^{iS_3}\right\}~.
\end{align}
It is very similar to the previous case: $S_4$ will play the role of $S_1$ and $S_3$ that of $S_2$.

We find that $S_4=-S_1$ and $S_3=-S_2$. Therefore the flow is the time-reversed of the previous one
\be
\mathfrak{S}'_2=-\mathfrak{S}'_{1}\,.
\ee

In the figures (\ref{fig8}) and (\ref{fig9}) below, we present the Bohmian trajectories corresponding to the wave solution $\Psi_2$.

\begin{figure}[H]
\begin{center}
\includegraphics[width=0.45\textwidth]{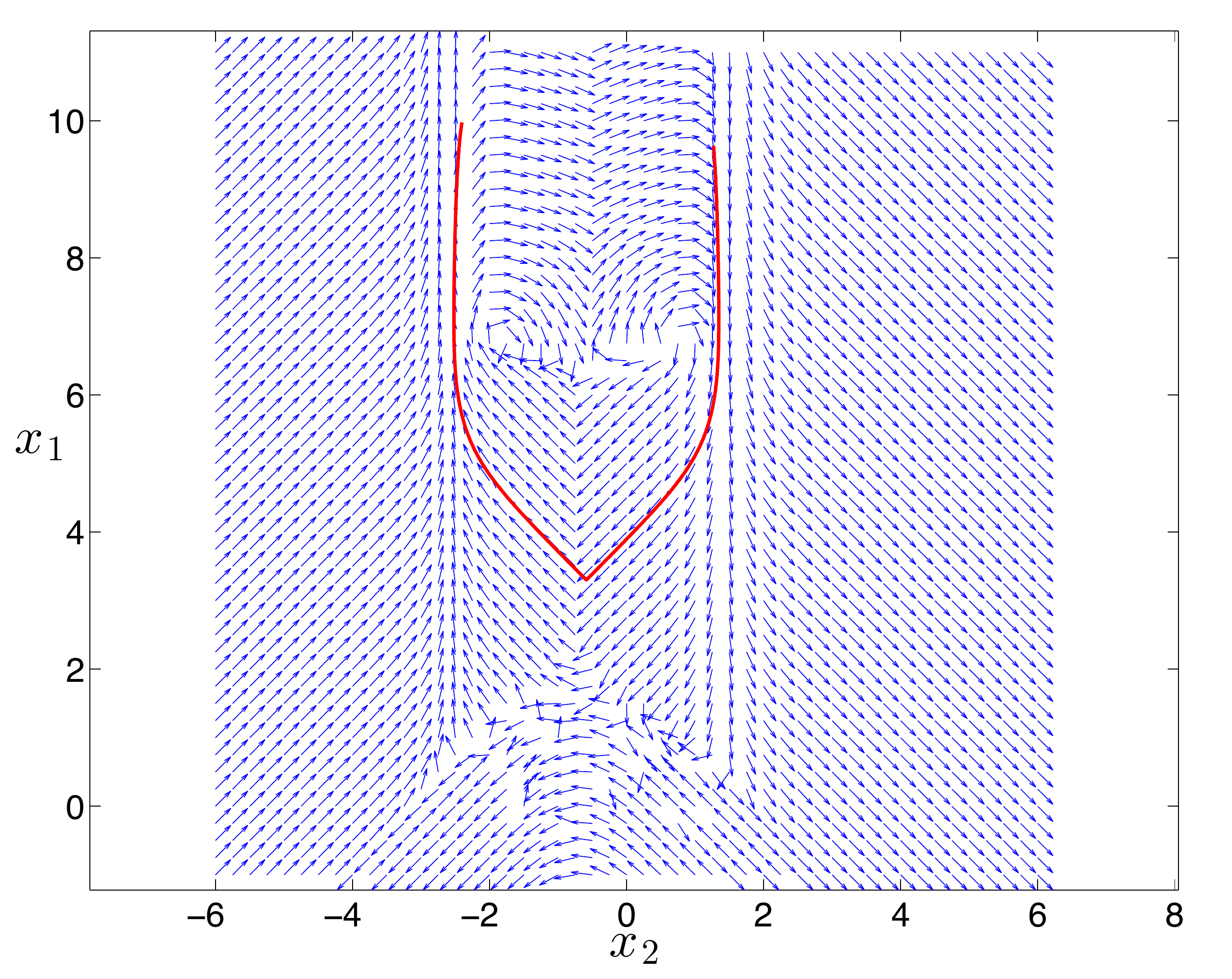}
\end{center}
\caption{Normed dBB flow for $\Psi_2$ (see (\ref{bigpsi2}) with $\sigma=0.1$, $k=10$ and $E=\sqrt{101}$), together with the dBB trajectory
(initial condition $(9.6362,1.2677)$, the final position of the trajectory plotted in Fig. (\ref{fig6}), and $t\in[0,10^{7}]$.
The dBB flow is the opposite of the one used for Fig. (\ref{fig6}).}
\label{fig8}
\end{figure}
\begin{figure}[H]
\begin{center}
\includegraphics[width=0.45\textwidth]{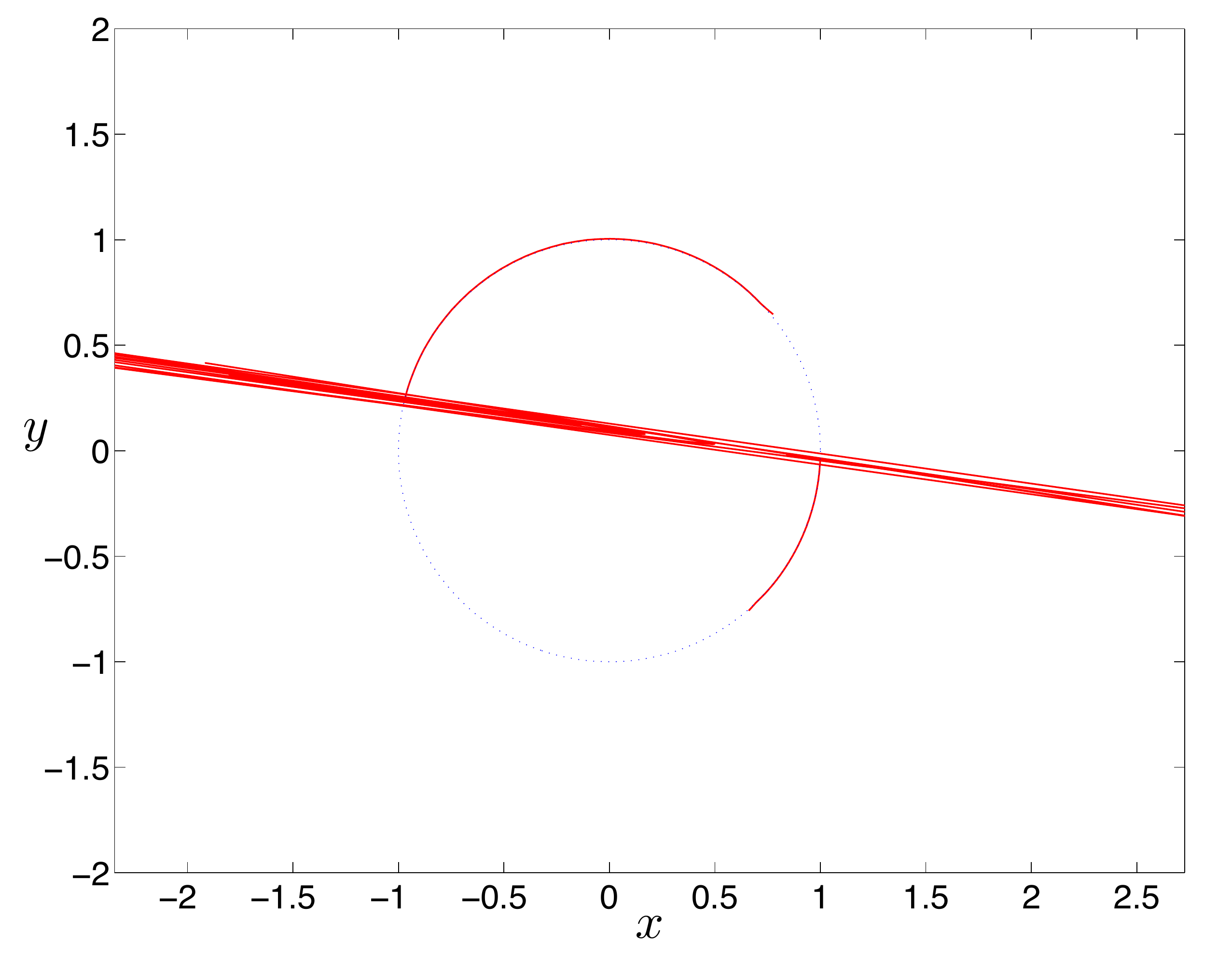}
\end{center}
\caption{Trajectory of Fig. (\ref{fig8}) visualized in the phase-space of Heard and Wands. The universe starts in the neighborhood
of $(\frac{1}{\sqrt{2}},-\frac{1}{\sqrt{2}})$ and ends ups in the neighborhood of $(\frac{1}{\sqrt{2}},\frac{1}{\sqrt{2}})$.
We have that $x\in[-46,195]$ and $y\in[-28,7]$.}
\label{fig9}
\end{figure}

\begin{figure}[H]
\begin{center}
\includegraphics[width=0.45\textwidth]{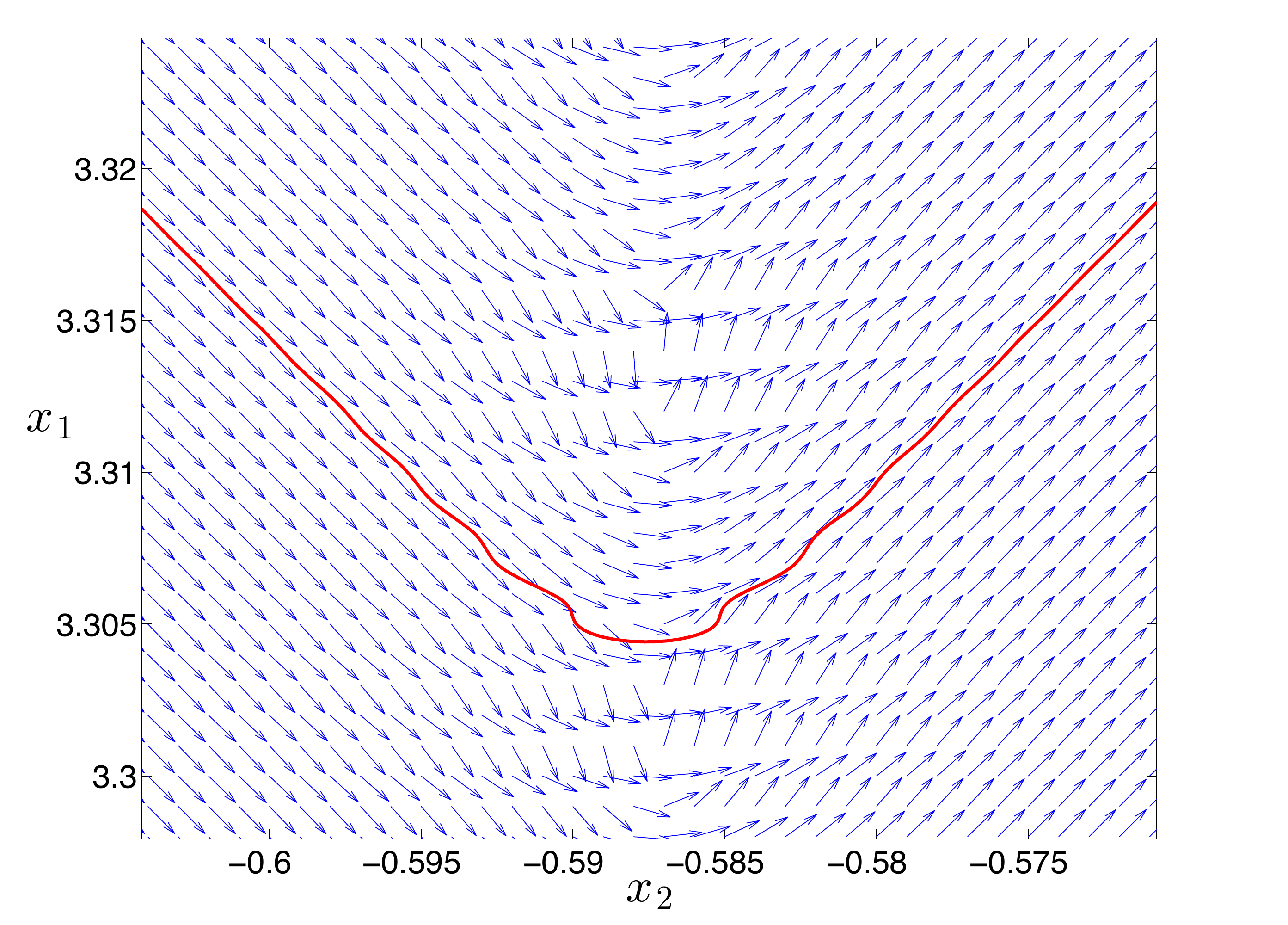}
\end{center}
\caption{Detail of the bounce corresponding to the trajectory of Fig. (\ref{fig6}).}
\label{figgoodbouncetraj4-detail}
\end{figure}

Fig. (\ref{figgoodbouncetraj4-detail}) shows the details of the bounce. Note that the trajectories can oscillate near the bounce,
a feature already noticed in Ref.~\cite{gaussian}.

\subsection{Bouncing trajectories in the Bessel-type basis}
\subsubsection{The de Broglie-Bohm flow}
Let us consider a wave-function of the type
\be\label{sol}
\Psi=\tilde{J}_{\nu}(\mathcal{E})e^{i k x_2}+c\tilde{Y}_{\nu}(\mathcal{E})e^{-i k x_2}~,
\ee
where $c=|c|e^{-i\theta}$.
The exact dBB velocity flow for the wave-function (\ref{sol}) is
\begin{align}
v_1=&\frac{|c|}{m_1|\Psi|^2}(\tilde{Y}_\nu(\mathcal{E}){\tilde{J}_\nu}'(\mathcal{E})
-{\tilde{Y}_\nu}'(\mathcal{E})\tilde{J}_\nu(\mathcal{E}))\sin(2 k x_2+\theta)~,&\\
v_2=&\frac{k}{m_2|\Psi|^2}(\tilde{J}_\nu^2(\mathcal{E})-|c|^2\tilde{Y}_\nu^2(\mathcal{E}))~,&
\end{align}
where $\nu=\frac{k}{\gamma}$, $\mathcal{E}=\frac{e^{\gamma x_1}}{\gamma}$, $m_2=e^{3\alpha}/(2 V_0)$
and $m_1=-m_2$.

For $\mathcal{E}\gg 1$, we have that
\begin{align}\label{flow1}
v_1\approx\frac{2}{\pi}\frac{\gamma|c|}{|m_1||\Psi|^2}\sin(2kx_2+\theta)\,,\nonumber\\
v_2\approx\frac{k}{m_2|\Psi|^2}\frac{2}{\pi \mathcal{E}}[1-(|c^2|+1)\sin^2(\mathcal{E})]~.
\end{align}
The sign of $v_1$ only depends on the position $x_2$. In the case $\theta=-\frac{\pi}{2}$, $\sin(2kx_2+\theta)\rightarrow-\cos(2 k x_2)$, and for
$\mathcal{E}\gg 1$ and $x_2\in]-\frac{\pi}{4 k}+n\frac{\pi}{k},\frac{\pi}{4 k}+n\frac{\pi}{k}[$, with $n\in\mathbb{Z}$, $v_1$ is negative. Elsewhere, for $\mathcal{E}\gg 1$, $v_1$ is positive. Furthermore, $|v_2|\ll|v_1|$. Therefore, if a universe starts in the vicinity of the attractor ($x_2=0$),
it will move towards the region $\mathcal{E}\approx 1$ (where the approximation fails). That part
of the trajectory would be almost parallel to the attractor
(due to the fact that $|v_2| \ll |v_1|$). Once the universe reaches the region where
the above approximation fails, the dBB flow will presumably become more complex and it is possible
that the universe could end up in a region of $x_2$ for which $v_1$ is positive at large
positive $\mathcal{E}$. In that case, that could lead to a bounce (without reaching any singularity).

Let us plot the actual normalized velocity field for a wave-function $\Psi_3$ obtained from (\ref{sol}) by putting $\theta=-\frac{\pi}{2}$, and for a wave-function $\Psi_4$ for which $\theta=0$.
\begin{figure}[H]
\begin{center}
\includegraphics[width=0.45\textwidth]{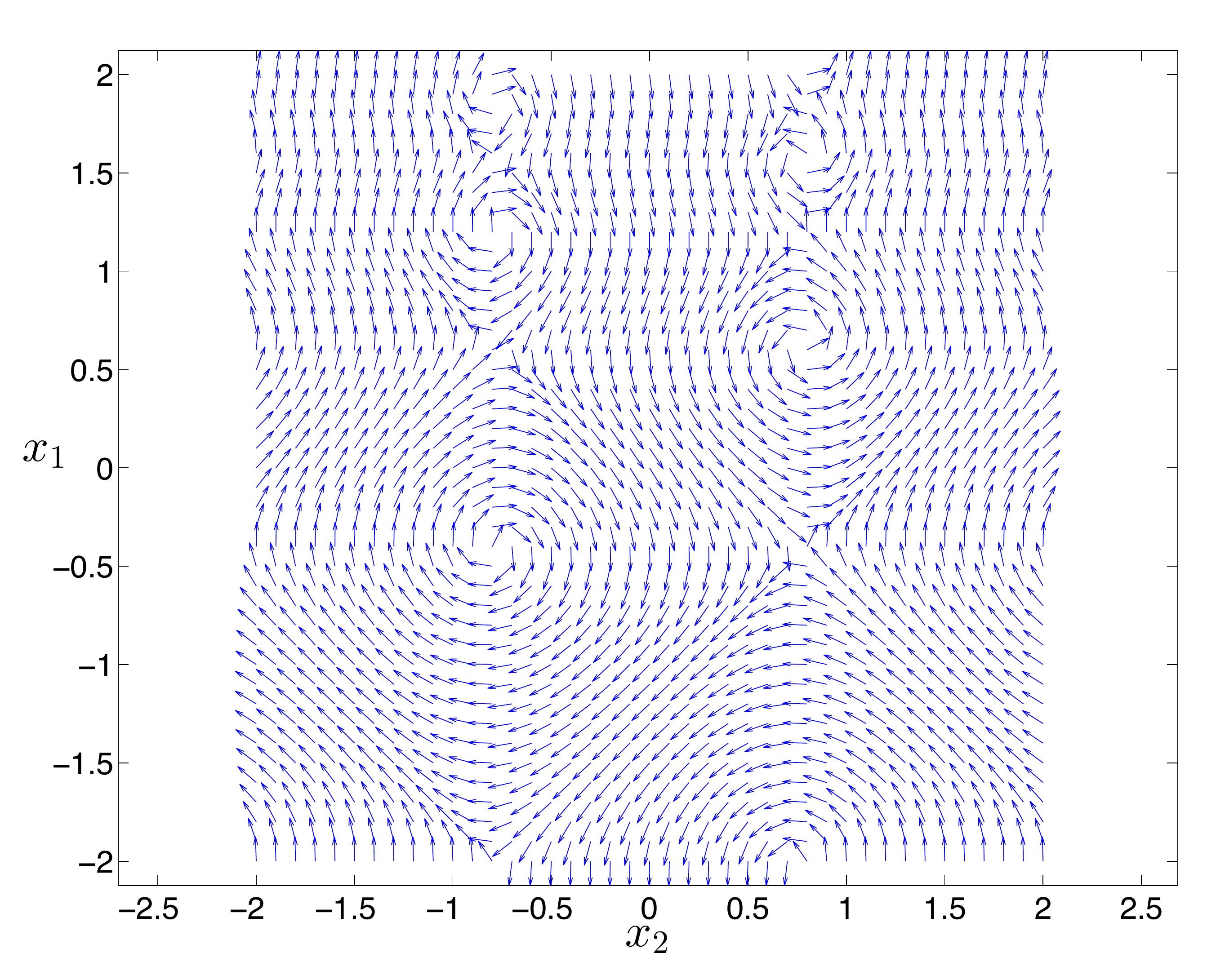}
\end{center}
\caption{The normalized dBB flow (\ref{nvf}) for the wave-function $\Psi_3$ (Eq.~(\ref{sol}) with $\theta=-\frac{\pi}{2}$). We use $ =1$, $k=1$, $V_0=\frac{9}{4}$, which leads to $\gamma=1$.}
\label{figdbbflow1}
\end{figure}
\begin{figure}[H]
\begin{center}
\includegraphics[width=0.45\textwidth]{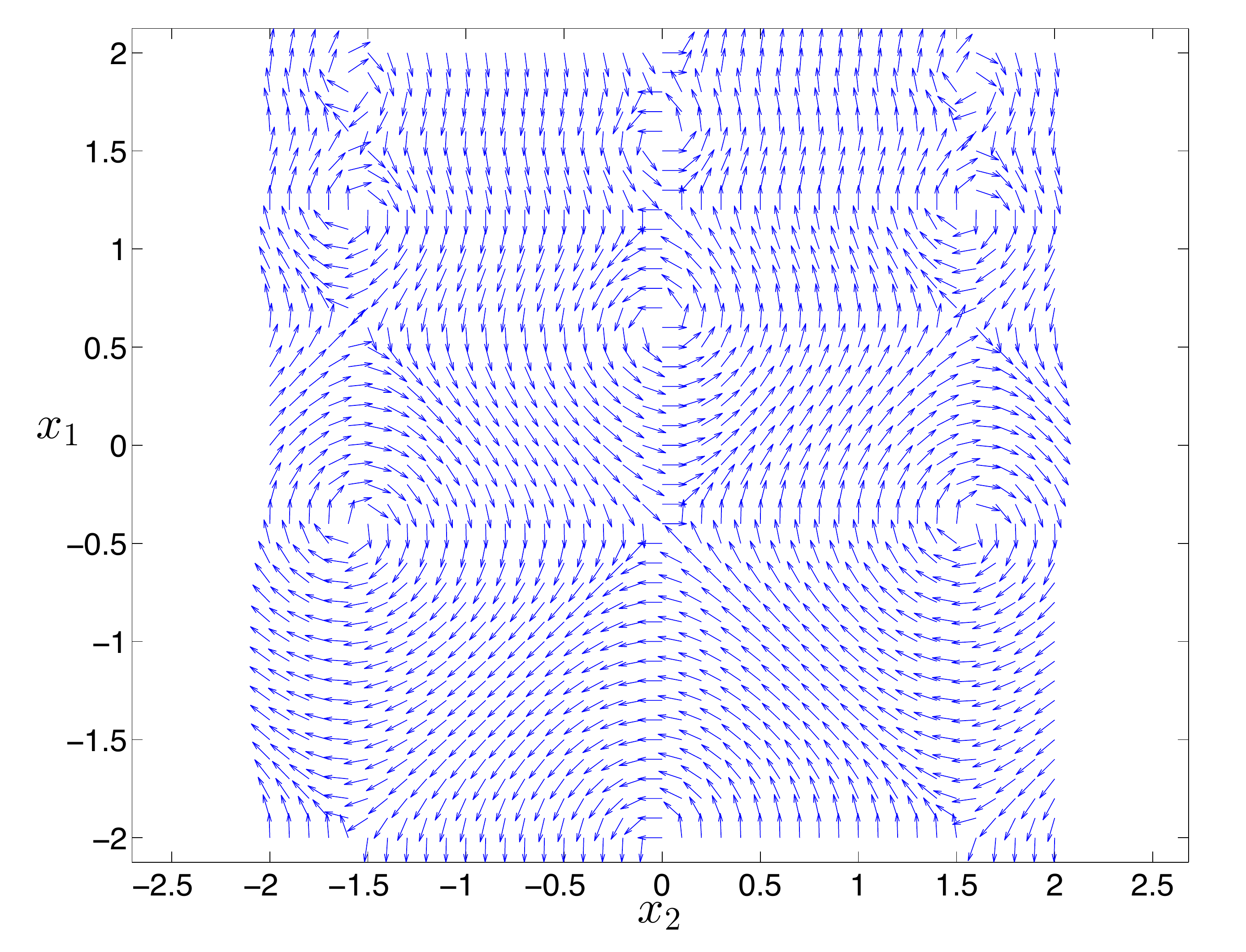}
\end{center}
\caption{The normalized dBB flow (\ref{nvf}) for the wave-function $\Psi_4$ (Eq. (\ref{sol}) with $\theta=0$). We use $ =1$, $k=1$, $V_0=\frac{9}{4}$, which leads to $\gamma=1$.}
\label{figdbbflow2}
\end{figure}
We see that both $\Psi_3$ and $\Psi_4$ allow bouncing trajectories.

For the trajectory equation, we have
\be
\frac{dx_1}{dx_2}=\frac{E'|c|}{k}
\frac{(\tilde{Y}'\tilde{J}-\tilde{Y}\tilde{J}')}{\tilde{J}^2-|c|^2\tilde{Y}^2}
\sin(2 k x_2 + \theta)~.
\ee
For $|c|=1$, we have that
\begin{align}
\frac{dx_1}{dx_2}=\frac{\mathcal{E}'}{k}
\frac{(\tilde{Y}'\tilde{J}-\tilde{Y}\tilde{J}')}{\tilde{J}^2-\tilde{Y}^2}
\sin(2 k x_2 + \theta)=\nonumber\\
\frac{\mathcal{E}'}{2k}\left(\frac{\tilde{J}'+\tilde{Y}'}{\tilde{J}+\tilde{Y}}
-\frac{\tilde{J}'-\tilde{Y}'}{\tilde{J}-\tilde{Y}}\right)
\sin(2 k x_2 + \theta)\,.
\end{align}
\subsubsection{Center and saddle points}
In the dBB velocity field, we only have alternating center and saddle points. As $x_1$ increases, the size of these center and saddle points diminish and they ultimately almost vanish.
\begin{figure}[H]
\begin{center}
\includegraphics[width=0.47\textwidth]{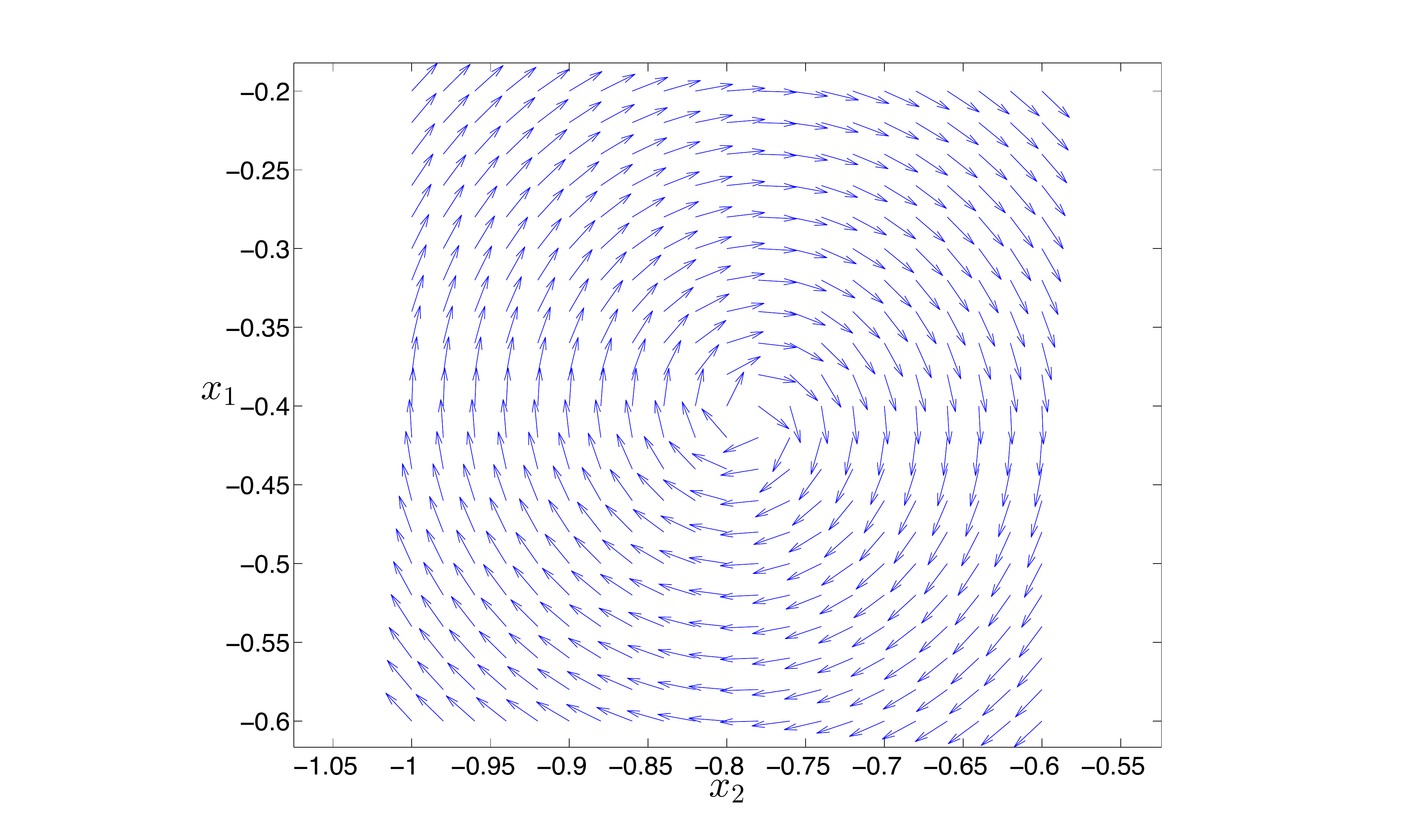}
\includegraphics[width=0.47\textwidth]{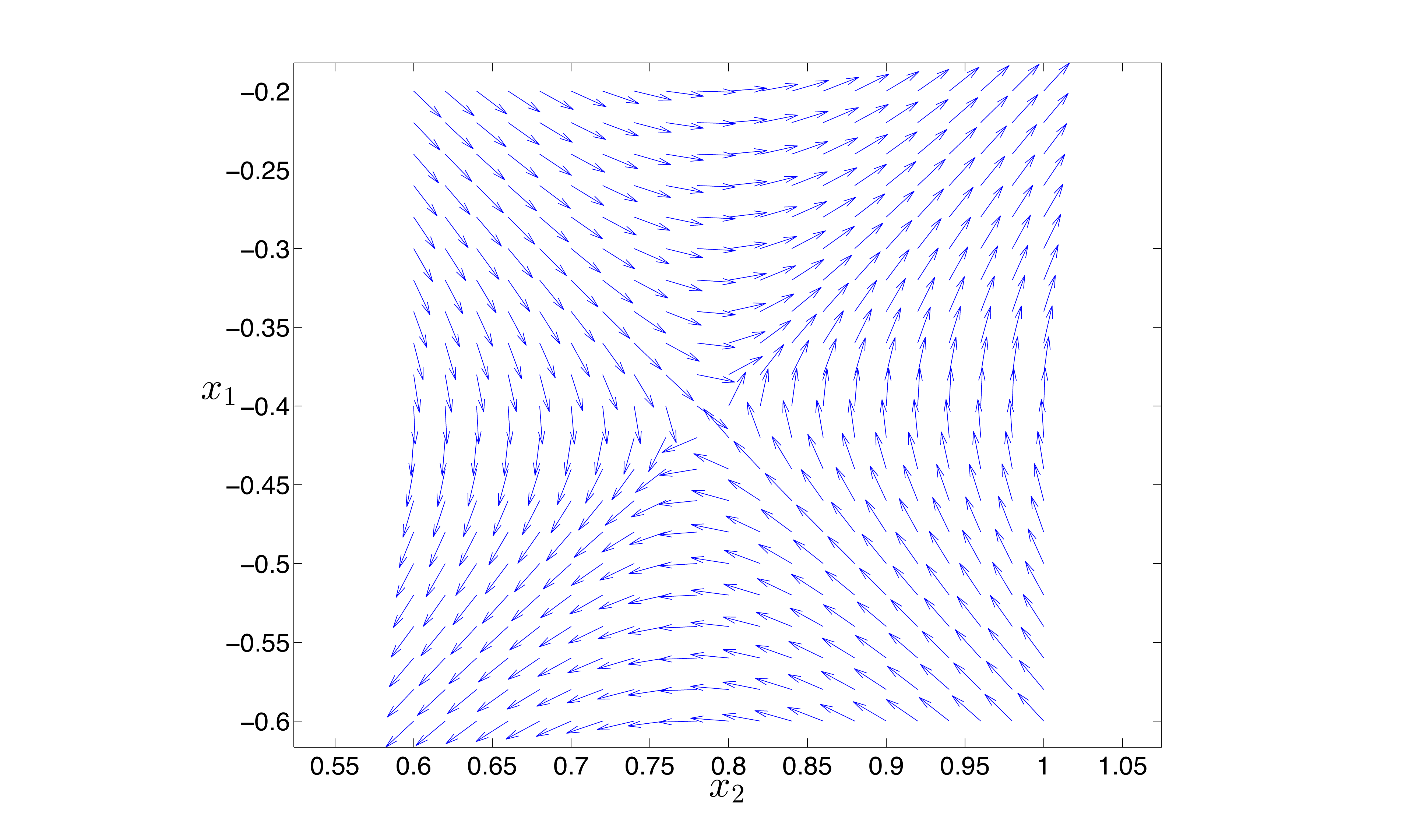}
\end{center}
\caption{A center and a saddle point. These are magnified regions of the normalized velocity field obtained for
$\Psi_3$ (see Fig. (\ref{figdbbflow1}))}
\label{figvorsad}
\end{figure}
We can deduce the positions of these vortices and saddle points by assuming
that the dBB velocity field is equal to zero there. Firstly, $v_1=0$ for $x_2=\frac{1}{2k}(\frac{\pi}{2}+n\pi-\theta)$.
Note that $v_1$ must be equal to zero in order for a center or saddle point to appear. One can see it from Figs. (\ref{figdbbflow1}) and (\ref{figdbbflow2}) (for which $k=1$).
In Fig. (\ref{figdbbflow1}), the center or saddle points are located at $x_2=-\frac{\pi}{4}$ or $x_2=\frac{\pi}{4}$.
In Fig. (\ref{figdbbflow2}), they are located at
$x_2=-\frac{\pi}{2}$, $x_2=0$ or $x_2=\frac{\pi}{2}$.
This fits with the formula $x_2=\frac{1}{2k}(\frac{\pi}{2}+n\pi-\theta)$.
For the second velocity field, $v_2=0$ for $x_1$ such that $\tilde{J}^2(\mathcal{E})=|c|^2\tilde{Y}^2(\mathcal{E})$.
We make a plot of $\tilde{J}^2(\mathcal{E})-\tilde{Y}^2(\mathcal{E})$ (for the case $|c|=1$).
\begin{figure}[H]
\begin{center}
\includegraphics[width=0.5\textwidth]{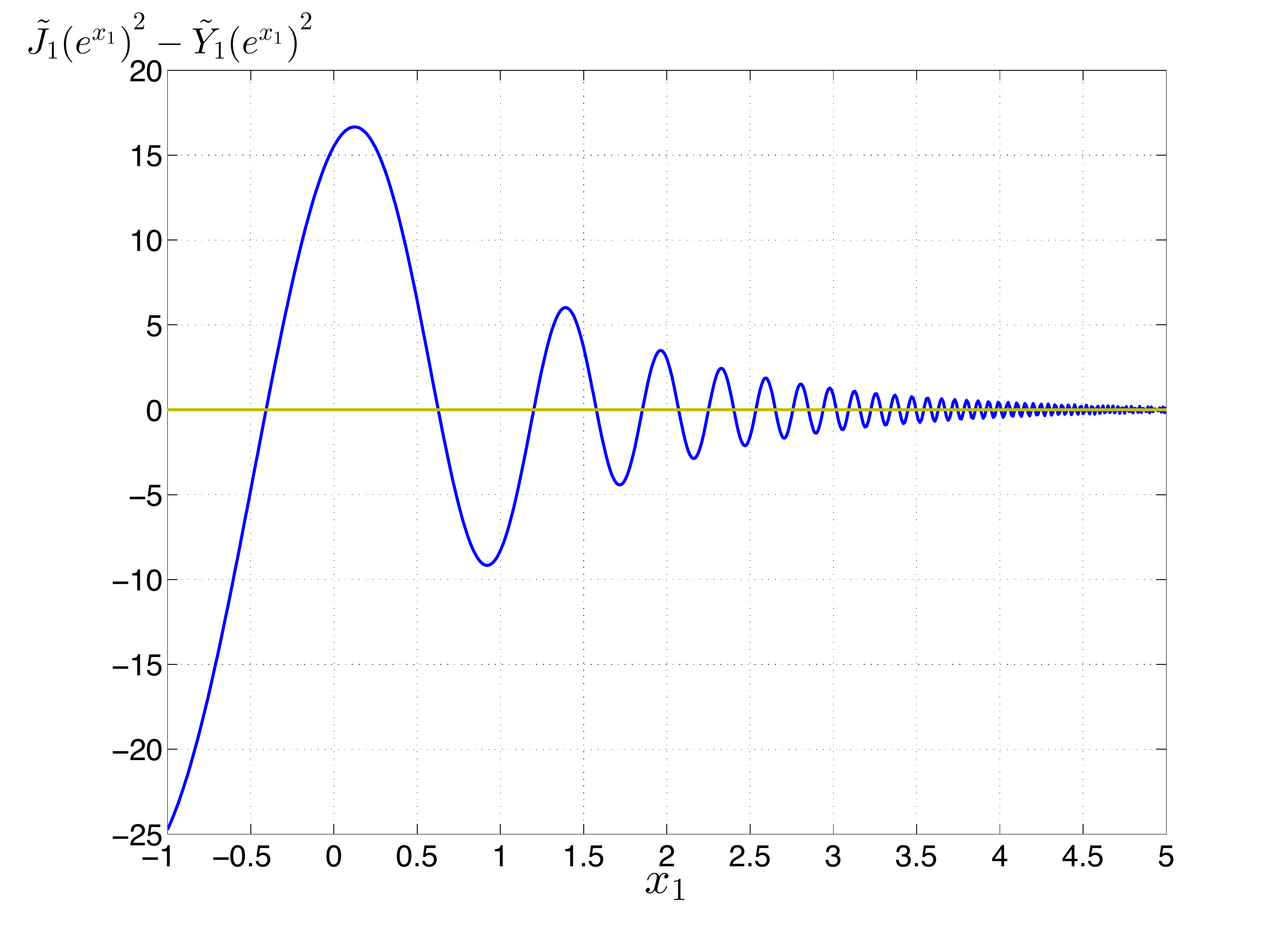}
\end{center}
\caption{Plot of $\tilde{J}^2_1(e^{x_1})-\tilde{Y}^2_1(e^{x_1}).$}
\end{figure}
The $x_1$-location of the vortices or saddle points correspond to the zeros of the above function.
\subsubsection{A universe starting on the classical attractor}
Let us consider the case of the wave function $\Psi_3$.
We consider the evolution of a universe starting on the classical attractor (at position $(5,0)$) and evolving from $t=0$ to $t=50$.
\begin{figure}[H]
\begin{center}
\includegraphics[width=0.5\textwidth]{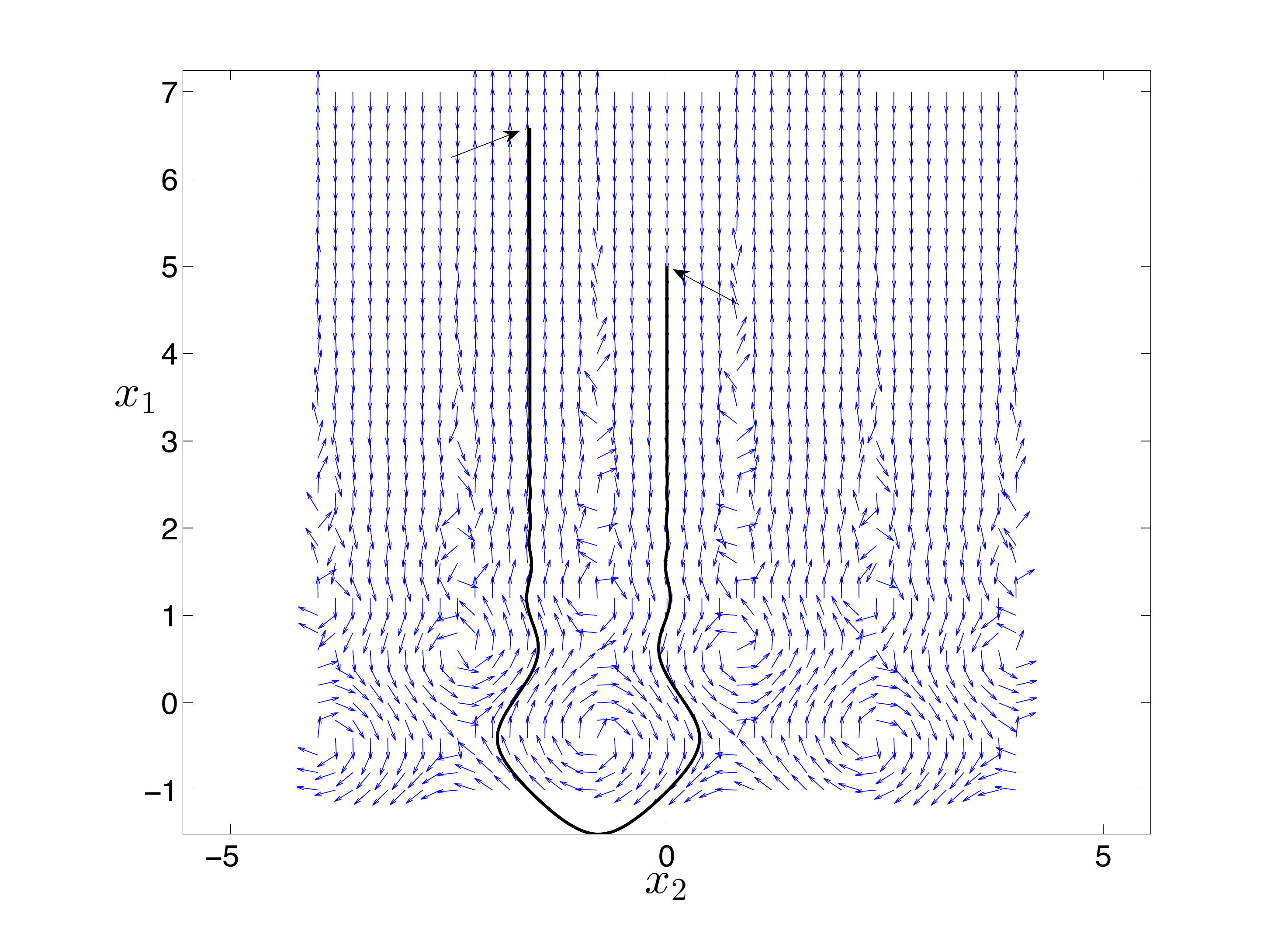}
\end{center}
\caption{Bouncing universe originating from the classical attractor.
The solution of the WDW equation is $\Psi_3$ (Eq.~(\ref{sol}) with $\theta=-\frac{\pi}{2}$).
The arrows point to the ending points of the trajectory. }
\label{figbouncetraj}
\end{figure}
We see that the universe undergoes a bounce.
We can also visualize this trajectory in the $(x,y)$
phase-space of Heard and Wands \cite{hewa} (see Fig. (\ref{xy1})).
\begin{figure}[H]
\begin{center}
\includegraphics[width=0.5\textwidth]{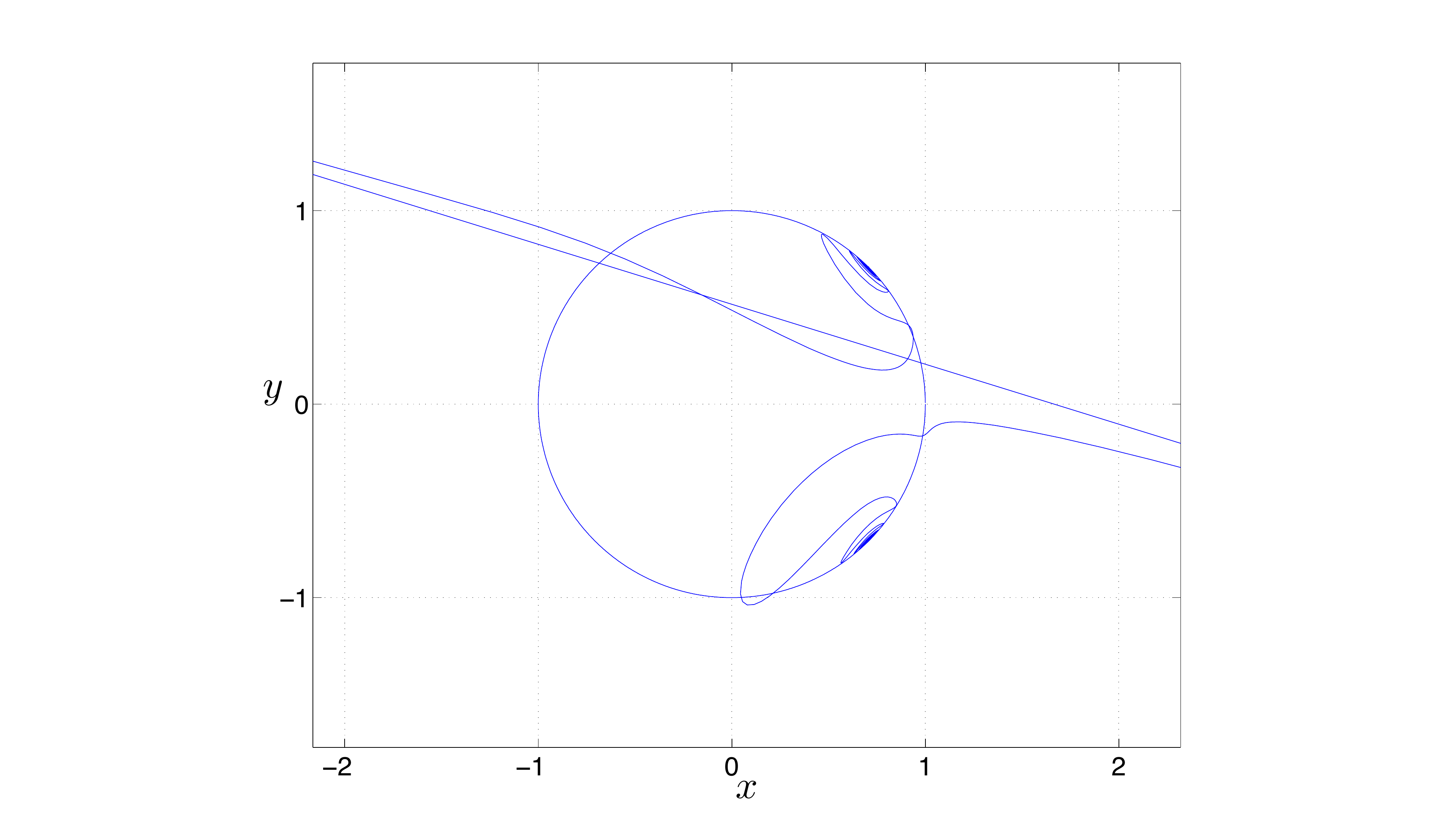}
\end{center}
\caption{Trajectory from Fig. (\ref{figbouncetraj}) in the phase-space of Heard and Wands.}
\label{xy1}
\end{figure}
We see that the universe starts on the classical circle (contracting phase, lower part) and ends up on the classical circle (expanding phase, upper part).
Hence, it starts at the dust repeller and ends ups at the dust attractor, but they leave the classical behavior and come back to it very soon and very late, respectively.
They do not pass through the dark energy and stiff matter phases.

Let us now see how the classical dust contraction is achieved in this solution.
$\Psi_3$ is a superposition of the basis functions
\be
\tilde{J}_{\nu}(\mathcal{E})e^{i k x_2}\textrm{ and }\quad \tilde{Y}_{\nu}(\mathcal{E})e^{i k x_2}\, .
\ee
When $\mathcal{E}\gg 1$, we have that $(\tilde{J}_{\nu}(\mathcal{E}))'\approx -\gamma \mathcal{E} \tilde{Y}_{\nu}(\mathcal{E})$ and $(\tilde{Y}_{\nu}(\mathcal{E}))'\approx \gamma \mathcal{E} \tilde{J}_{\nu}(\mathcal{E})$.
Therefore, there is an extra factor $\mathcal{E}$ in $\partial_1\Psi_3$ with respect to $\partial_2 \Psi_3$. This explains why the dBB velocity is almost parallel to the $x_1$ axis when $\mathcal{E}\gg 1$.
This in turn means that the universe moves on line $x_2=constant$ at $\mathcal{E}\gg 1$, which means that $d\phi-d\alpha/\sqrt{2}=0$ and $\frac{d\phi}{d\alpha}=\frac{1}{\sqrt{2}}$.
The x coordinate of Heard and Wands is $\frac{d\phi}{d\alpha}$ and hence it is equal to $1/\sqrt{2}$.

The $y$ coordinate is defined as
\be
y=\sqrt{2|V|}/{\dot{\alpha}}~.
\ee
We have that $\alpha=(x_1+x_2/\sqrt{2}) /\sqrt{V_0}$. Therefore, for $\mathcal{E}\gg 1$, we have that $\dot{\alpha}\approx\frac{ \dot{x}_1}{\sqrt{V_0}}$ and
\be
y=\frac{\sqrt{2} V_0 e^{-\frac{3}{\sqrt{2}}\phi}}{  \dot{x}_1}~.
\ee
This quantity must be constant for a given $x_2$, and for $\mathcal{E}\gg 1$. Actually, it is not easy to enforce this condition. Let us see how it works with our example.
We have that
\be
\dot{x}_1=\frac{2}{\pi}\frac{\gamma}{|m_1||\Psi_3|^2}\sin(2 k x_2 +\theta)~,
\ee
with
\be
|\Psi_3|^2=\tilde{J}^2(\mathcal{E})+\tilde{Y}^2(\mathcal{E})+2\tilde{J}(\mathcal{E})\tilde{Y}(\mathcal{E})\cos (2 k x_2 +\theta)
\ee
and $|m_1|= e^{3\alpha}/(2 V_0)$. When the sine is equal to $-1$ and the cosine is equal to $0$, we have that $\dot{x}_1=-\frac{\mathcal{E}}{|m_1|}$ and
\begin{align}
y=\frac{\sqrt{2} V_0 e^{-\frac{3}{\sqrt{2}}\phi}(|m_1|)}{  (-\mathcal{E})}
=\frac{\sqrt{2} V_0 e^{-\frac{3}{\sqrt{2}}\phi}(e^{3\alpha})}{  2 V_0(-\mathcal{E})}
\nonumber\\=-\frac{1}{\sqrt{2} }=-\frac{1}{\sqrt{2}} .
\end{align}
Hence the classical dust contraction is achieved only for specific values of $x_2$.

Despite the fact that the trajectory has the correct starting and ending points, it quickly departs from the classical one. This is not the
case of the solutions coming from the semi-classical approximation: they leave the classical
evolution only near the stiff matter behavior, hence they are physically preferable as long as they allow a classical dark energy phase.

\begin{figure}[H]
\begin{center}
\includegraphics[width=0.5\textwidth]{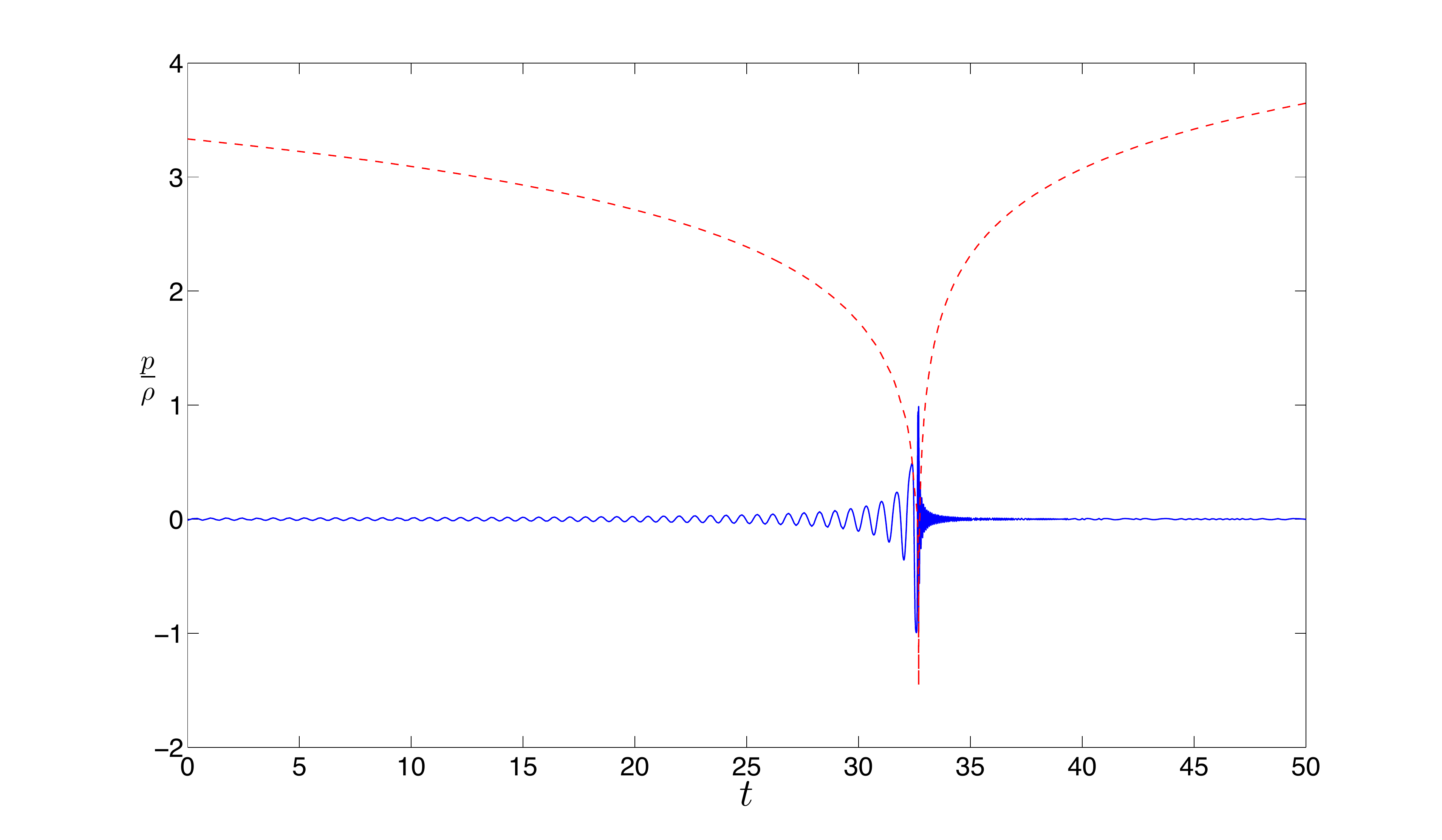}
\end{center}
\caption{Evolution of $\frac{p}{\rho}$ for a universe starting on the classical attractor (position $(5,0)$ at $t=0$). The dotted line corresponds to $\alpha(t)$.
The value $t_B$ for which $\alpha(t)$ is minimal corresponds, in this case, to the bounce.}
\label{figpoverrho}
\end{figure}

We plot in Fig. (\ref{figpoverrho}) the evolution of $p/{\rho}$, which is given by (\ref{ratio}).
Around the bounce, we have an oscillation between two behaviors: kinetic energy domination, and potential energy domination.

\section{Conclusion}

Scalar fields with exponential potential were investigated in the context of bouncing models, where the bounce happens due to quantum cosmological effects. The corresponding Wheeler-DeWitt equation was solved, and interpreted according to the de Broglie-Bohm quantum theory. The quantum trajectories describing the scale factor evolution were calculated, and the space of solutions was explored in its full generality. It was shown that there are two types of bouncing solutions. The first one comes from dust contraction but soon quantum effects become important and it makes a bounce. Classical behavior is recovered only at the expanding phase, near dust evolution again. The more interesting bouncing solution is the one where the classical dynamics remains valid up to stiff matter behavior. In this region, quantum effects become important and the bounce takes place. These solutions must have one and only one dark energy phase, either occurring in the contracting era or in the expanding era. They are necessarily asymmetric. Taking the realistic situation where the dark energy phase happens in the expanding era, one has the picture of a universe realizing a classical dust contraction from very large scales, the initial repeller of the model, moving to a stiff matter contraction near the singularity, which is avoided due to the quantum bounce. The universe is then launched to a stiff matter expanding phase, which then moves to a dark energy era, finally returning to the dust expanding phase, the final attractor of the model. Hence, such an exponential potential scalar field, in one stroke, can not only describe the matter contracting phase of a nonsingular bouncing model, necessary to give an almost scale invariant spectrum of scalar cosmological perturbations, but it can also model a transient expanding dark energy phase. Furthermore, as the universe is necessarily dust dominated in the far past, usual adiabatic vacuum initial conditions can be easily imposed in this era. Hence, this is a cosmological model where the presence of dark energy in the universe does not turn problematic the usual initial conditions prescription for cosmological perturbations in bouncing models and, consequently, it is able to yield a well posed problem to calculate the observed spectrum and amplitude of scalar cosmological perturbations in bouncing models with dark energy~\cite{beatriz}. Note that usual investigations in LQC do not explore this interesting richer evolution of the background in the classical domain allowed by the exponential potential. This is something to be better investigated in this quantization framework.

The evolution of scalar and tensor cosmological perturbations were recently investigated in detail in such models, see Ref.~\cite{annapaula}. There it was found, through analytical arguments and detailed numerical calculations, including through the bounce itself, that we can obtain almost scale invariant spectrum of scalar perturbations with the right amplitude. The qualitative mechanism is almost the same as in usual single field bouncing cosmologies. However, one distinct feature of this quantum bounce was the enhancement of scalar perturbations over tensor perturbations, something which does not take place in classical bounces, but which was already noticed in Ref.~\cite{cai-ewing1} within their particular quantum bounce. In our scenarios, however, this enhancement can be quite significative, and the reasons for that are discussed in Ref.~\cite{annapaula}. Such property solves the problem concerning the high tensor-to-scalar ratio generally present in single field bouncing models with canonical kinetic term (and hence with $c_s=1$)\cite{lqc3}.
 
Of course this is still a toy model, but the fact that such a simple model already contains so many good features is a good motivation to search for more realistic extensions. One possibility is to consider the classical extension analyzed in Ref.~\cite{extension}, which takes one hydrodynamical fluid with $p=w\rho$, $w=$const. besides the scalar field with exponential potential, and to study its canonical quantization and quantum trajectories. This is one of the subjects of our future research along these lines.
\begin{acknowledgments}
S.C. and N.P.N. would like to thank CNPq of Brazil for financial support.
N.P.N. would like to thank Diego Pantoja and Daniel Wagner for many valuable calculations and discussions in the beginning of this research.
\end{acknowledgments}
\appendix
\section{A relation between the two basis of functions}
In \cite{graryz}, we find (p. 482, formula 3.876.1 and 3.876.2) that
\begin{align}
\int_{0}^{\infty}\frac{\sin(p\sqrt{x^2+a^2})}{\sqrt{x^2+a^2}}\cos(bx)=
\frac{\pi}{2}J_0(a\sqrt{p^2-b^2})\,,\\
\int_{0}^{\infty}\frac{\cos(p\sqrt{x^2+a^2})}{\sqrt{x^2+a^2}}\cos(bx)=
-\frac{\pi}{2}Y_0(a\sqrt{p^2-b^2})\,,
\end{align}
for $0<b<p$.
With the following replacements, $x\rightarrow k$, $a\rightarrow 1$, $p\rightarrow u$ and
$b\rightarrow |v|$, we have that
\begin{align}
\int_{0}^{\infty}\frac{\sin(u\sqrt{k^2+1})}{\sqrt{k^2+1}}\cos(|v|k)=
\frac{\pi}{2}J_0(\sqrt{u^2-v^2})\,,\\
\int_{0}^{\infty}\frac{\cos(u\sqrt{k^2+1})}{\sqrt{k^2+1}}\cos(|v|k)=
-\frac{\pi}{2}Y_0(\sqrt{u^2-v^2})\,,
\end{align}
for $0<v<u$.
In the $(u,v)$ coordinate system, $u^2-v^2\geq 0$ but $v$ can be negative. However, since
\begin{align}
\Psi=&\int_{-\infty}^{\infty} e^{i(\pm u\sqrt{k^2+1}+kv)}\frac{1}{\sqrt{k^2+1}}dk&\nonumber\\
=&2\int_{0}^{\infty}\biggl[\frac{\cos(\sqrt{k^2+1}u)\pm i\sin(\sqrt{k^2+1}u)}{\sqrt{k^2+1}}\biggr]\cos(vk)dk~,&
\end{align}
v only enters as a factor in the argument of the cosine.
Therefore, we can simply assume that $v=|v|$, and we have that
\begin{align}
\Psi(u,v)=-\pi Y_0(\sqrt{u^2-v^2})\pm i\pi J_0(\sqrt{u^2-v^2})=\nonumber\\
-\pi(Y_0(\sqrt{u^2-v^2})\mp i J_0(\sqrt{u^2-v^2}))
~.
\end{align}
Thus the pos-E solution is given by $Y_0+i J_0$, while the neg-E solution is given by $Y_0-i J_0$.
Also we have that $\sqrt{u^2-v^2}=e^{\gamma x_1}\gamma^{-1}$.

It is a link between the KG-type and the Bessel-type solutions: we build a packet of KG-type solutions, and
we end up with the $J_{0}$ and $Y_{0}$ solutions.

\bibliographystyle{plain}

\end{document}